\documentclass[prb,twocolumn,superscriptaddress]{revtex4-1}
\usepackage{amsmath,amssymb,mathrsfs}
\usepackage{graphicx}
\usepackage{epstopdf}
\usepackage{color}

\begin{document}

\title{Universality of Heisenberg-Ising chain in external fields}

\author{Haiyuan Zou}
\affiliation{Tsung-Dao Lee Institute \& School of Physics and Astronomy, Shanghai Jiao Tong University, Shanghai 200240, China}

\author{Rong Yu}
\email{rong.yu@ruc.edu.cn}
\affiliation{Department of Physics and Beijing Key Laboratory of Opto-electronic Functional Materials \& Micro-nano Devices, Renmin University of China, Beijing, 100872, China}

\author{Jianda Wu}
\email{wujd@sjtu.edu.cn}
\affiliation{Tsung-Dao Lee Institute \& School of Physics and Astronomy, Shanghai Jiao Tong University, Shanghai 200240, China}

\begin{abstract}
Motivated by the recent surge of transverse-field experiments on
quasi-one-dimensional antiferromagnets Sr(Ba)Co$_2$V$_2$O$_8$,
we investigate the quantum phase transition in a Heisenberg-Ising chain under
a combination of two in-plane inter-perpendicular transverse fields
and a four-period longitudinal field,
where the in-plane transverse field is either uniform or staggered.
We show that the model can be unitary mapped to
the one-dimensional transverse-field Ising model (1DTFIM) when the 
$x$ and $y$ components of the spin interaction and the four-period field are absent. When these two terms
are present, following both 
analytical and numerical efforts, we demonstrate
that the system undergoes a second-order quantum phase
transition with increasing transverse fields, where the critical exponents
as well as the central charge fall into the universality of 1DTFIM. Our results
naturally identify the 1DTFIM universality
of 1D quantum phase transitions observed in the
existed experiments 
in Sr(Ba)Co$_2$V$_2$O$_8$ with transverse
field applied along either [100] or [110] direction.
Upon varying the tuning parameters a critical surface with 1DTFIM universality is
determined and silhouetted to exhibit
the general presence of the universality in a much wider scope of
models than conventional understanding.
Thus our results provide a broad
guiding framework to facilitate the
experimental realization of 1DTFIM universality in real materials.
\end{abstract}

\maketitle
\section{Introduction}
\label{intro}
Quantum phase transition arises when the ground state energy
of a many-body system encounters non-analyticity
upon non-thermal parameter tuning~\cite{SpecialIssue2010,sachdev2011}.
A quantum critical point (QCP) pops up when the 
transition
is continuous.
Near a QCP, exotic behaviors which have no
counterparts in classical phase transitions
emerge due to 
the intrinsic driven power of quantum fluctuations.
These include the non-Fermi-liquid behavior and unconventional
superconductivity in a variety of strongly correlated electron
systems~\cite{SpecialIssue2010,Si2001,Schroder2000, Coleman2005,Si_Science2010,Schuberth2016,Wu2016b},
as well as the peculiar spin dynamics in
one-dimensional
quantum magnets~\cite{Kinross2014,WuPRL2014}.

The one-dimensional transverse-field Ising model (1DTFIM) serves as 
a paradigmatic quantum
spin model that exhibits a continuous quantum phase transition with
rich quantum critical behaviors \cite{Niemeijer1967,Pfeuty1970,McCoy1971,Suzuki1971,Suzuki1976,
Jullien1978,Chakravarty2005,WuPRB2018,WangPRL2018},
where the celebrated conformal invariance
emerges near its QCP \cite{Belavin1984}.
Although the quantum criticality of the 1DTFIM has
been theoretically 
studied in great detail during past decades,
it remains rare
and tough to experimentally detect this 1D quantum criticality: 
On one hand, spin interactions beyond the standard 1DTFIM are usually non-negligible in real materials;
on the other hand, the 3D
ordering at finite temperatures 
may mask the
1D quantum critical point and the associated
critical behavior \cite{ColdeaE8Science2010}.

Among the potential 
materials,
the widely investigated quasi-one-dimensional
effective spin-$1/2$ antiferromagnetic Heisenberg-Ising 
screw chain compounds SrCo$_2$V$_2$O$_8$
(SCVO)~\cite{HePRB2006SCVO,BeraPRB2017,BeraPRB2014,WangNature2018,NP2018Faure} and
BaCo$_2$V$_2$O$_8$ (BCVO)~\cite{HePRB2005BCVO,KimuraPRL2007,LakePRL2013,KlanjsekPRB2015,KimuraJPSJ2013}
are encouraging candidates to access the 1D QCP. 
Recent series transverse-field experiments on these
two materials confirm the existence of a novel 1D QCP residing
outside the dome of the 3D N\'{e}el ordered phase,
which further reveals
promising features of quantum criticality of 1DTFIM universality
around the 1D QCP\cite{WangPRL2018,YuRecent}.
However, the underlying effective 
model 
for understanding this 1D QPT is complicated. Owing to the screw chain structure, the applied transverse
field induces an in-plane staggered field perpendicular to it 
and a four-period field along the crystalline $c$ axis. This gives rise 
to a 
Heisenberg-Ising model 
under various external fields [{\it cf.} Eq.~(\ref{eq:model1})].

This superficially complicated 
model looks completely
different from the 
original 1DTFIM.
It raises a question about the nature of
the 
QCP, if there exists one,
in the model [{\it cf.} Eq.~(\ref{eq:model1})].
To clarify this issue, we systematically carry out theoretical investigation
on the quantum phase transition and the possible universality class near
the QCP of the effective model via determining various
critical exponents as well as the central charge 
associated with the conformal invariance.

Application of a transverse magnetic field suppresses the long-range antiferromagnetic (AF) order
and induces an order-disorder phase transition. 
As a consequence of the anisotropy of the crystal structure, the value of the critical transverse field varies with the field direction. 
For example, when the transverse field is applied along the $[100]$ direction,
a staggered transverse field 
along the $[010]$ direction and a four-period-staggered field 
along the $[001]$ direction ($c$ axis) are induced~\cite{KimuraJPSJ2013}. The induced fields shift the quantum critical
point $H_c$ to $\sim$10 T for BCVO~\cite{NiesenPRB2013} and $\sim$7 T for SCVO~\cite{WangPRB2016},
significantly lower than the $H_c$ values when 
the field is applied along the $[110]$ direction~\cite{WangPRL2018}.
Given that the interchain exchange coupling is much 
weaker than the intrachain coupling along the
easy-axis ($c$-axis) 
in these compounds, these order-disorder phase transitions 
exhibit sharp features and can be described by an effective 1D spin-$1/2$ XXZ model with external fields.

In this paper, we study 
this effective
1D model [{\it cf.} Eq.~(\ref{eq:model1})] with
corresponding parameters following the existed
materials of SCVO and BCVO.
By using infinite time-evolving block decimation
(iTEBD) method~\cite{iTEBD1,iTEBD2}, 
we calculate the staggered magnetization, the entanglement entropy, and the spin-spin
correlation function with fields 
to locate the QCP. Meanwhile, the scaling analysis on these quantities allows us to extract various critical exponents and the central charge
to characterize the nature of the quantum phase transition.
The phase transition is found to fall into the 1DTFIM
universality for the transverse field 
applied along the [100] (or equivalently [010])
direction in real experiments~\cite{YuRecent}.
We further demonstrate that the 
induced $c$-axis
four-period-staggered field is irrelevant for the 
universality of the transition.
After pinning down the nature of the 1D QCP in the existed
transverse-field experiments on SCVO and BCVO,
more general situations of the model are further 
analyzed in detail
by relaxing the allowed tuning fields in the effective model.
With varying the spin anisotropy factor $\epsilon$ and 
the transverse components of the $g$ factor $g_x$ and $g_y$, 
a critical surface falling into 1DTFIM universality is determined
and silhouetted
in the parameter space.
The numerical calculation is subsequently followed by a mean field analysis, 
which qualitatively
confirms the iTEBD results for small $\epsilon$. The well-established
critical surface immediately leads to promising prediction
that a line of 1DTFIM critical point should be directly observed
when the transverse field is gradually rotated from [100] to [110] direction
in 
either SCVO or BCVO.

The rest of this paper is organized as follows. Sec.~\ref{model1} 
presents the effective model for 
SCVO and BCVO, and shows that 
the quantum phase transition 
induced by tuning the applied transverse field 
belongs to the 1DTFIM universality class.
Sec.~\ref{generalized} studies the model in more general situations. 
We obtain the ground-state phase diagram of this generalized model by iTEBD calculations. These numerical results are further confirmed 
by a mean-field analysis.
Sec.~\ref{experiment} provides
a guideline for experiments based on the general theoretical results.
We draw conclusion in Sec.~\ref{conclusion}. 

\section{The effective model}
\label{model1}

For the material of SCVO or BCVO when the transverse field is applied along
the $[100]$ direction, the corresponding effective spin-$1/2$ Heisenberg-Ising
XXZ Hamiltonian reads as 
\begin{eqnarray}
\label{eq:model1}
\nonumber
&&H=\sum_iJ[\epsilon(S^x_iS^x_{i+1}+S^y_iS^y_{i+1})+S^z_iS^z_{i+1}]- \\
 &&g\sum_i[S^x_i+h_y(-1)^iS^y_i+h_z\cos(\pi\frac{2i-1}{4})S^z_i],
\end{eqnarray}
where $\vec{S}=\vec{\sigma}/2$, and $\vec{\sigma}$ refer to the three Pauli matrices.
Take the SCVO system for example, $J=1.67$ meV, and the anisotropic factor $\epsilon\sim 0.5$ are estimated~\cite{WangPRB2015}.
$g=\mu_BG_xh$ with Bohr magneton $\mu_B$, the gyromagnetic ratio $G_x=2.79$,\cite{WangPRB2015}
and $h$ is the applied $a$-axis magnetic field. 
The ratio of the induced staggered fields to $g$ are set as $h_{y(z)}=0.4(0.14)$~\cite{KimuraJPSJ2013}. In our calculation,
we set the energy unit $J=1$ in the model, and take the anisotropic factor $\epsilon=0.47$. 
In this way, $g$ 
is the only tuning parameter in the model. 

\begin{figure*}
\includegraphics[width=0.35\textwidth 
]{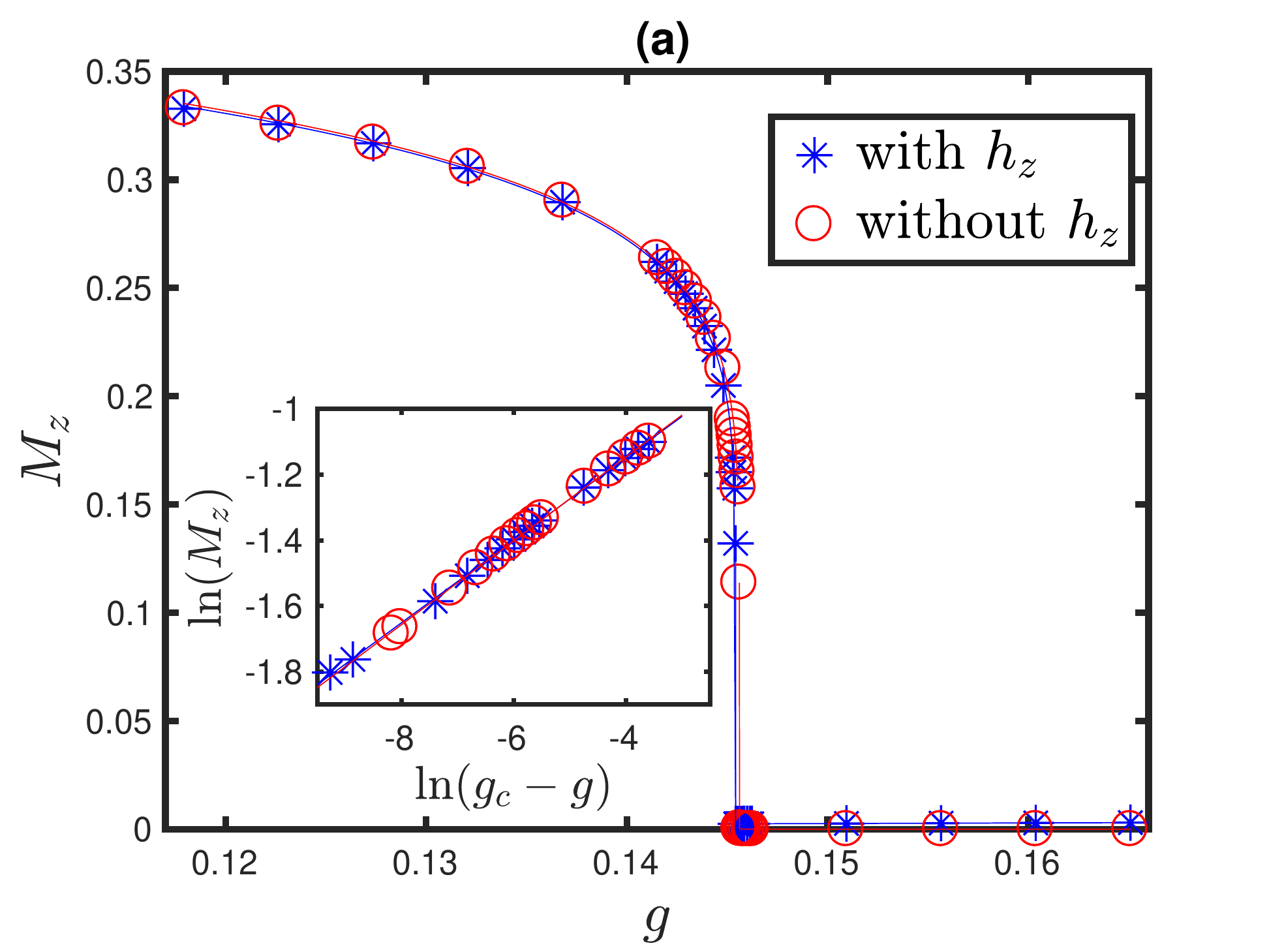}
\includegraphics[width=0.35\textwidth 
]{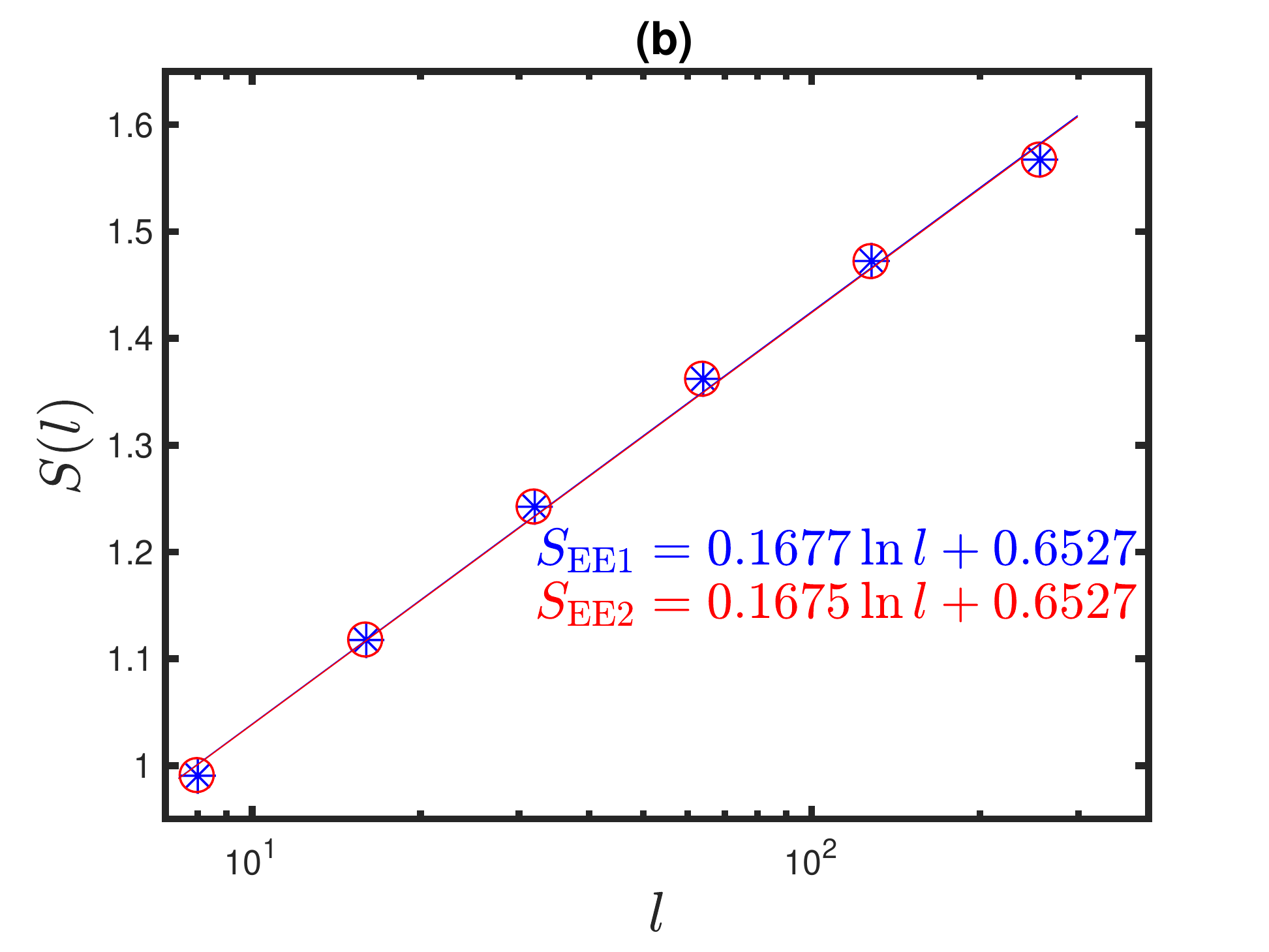}
\includegraphics[width=0.35\textwidth 
]{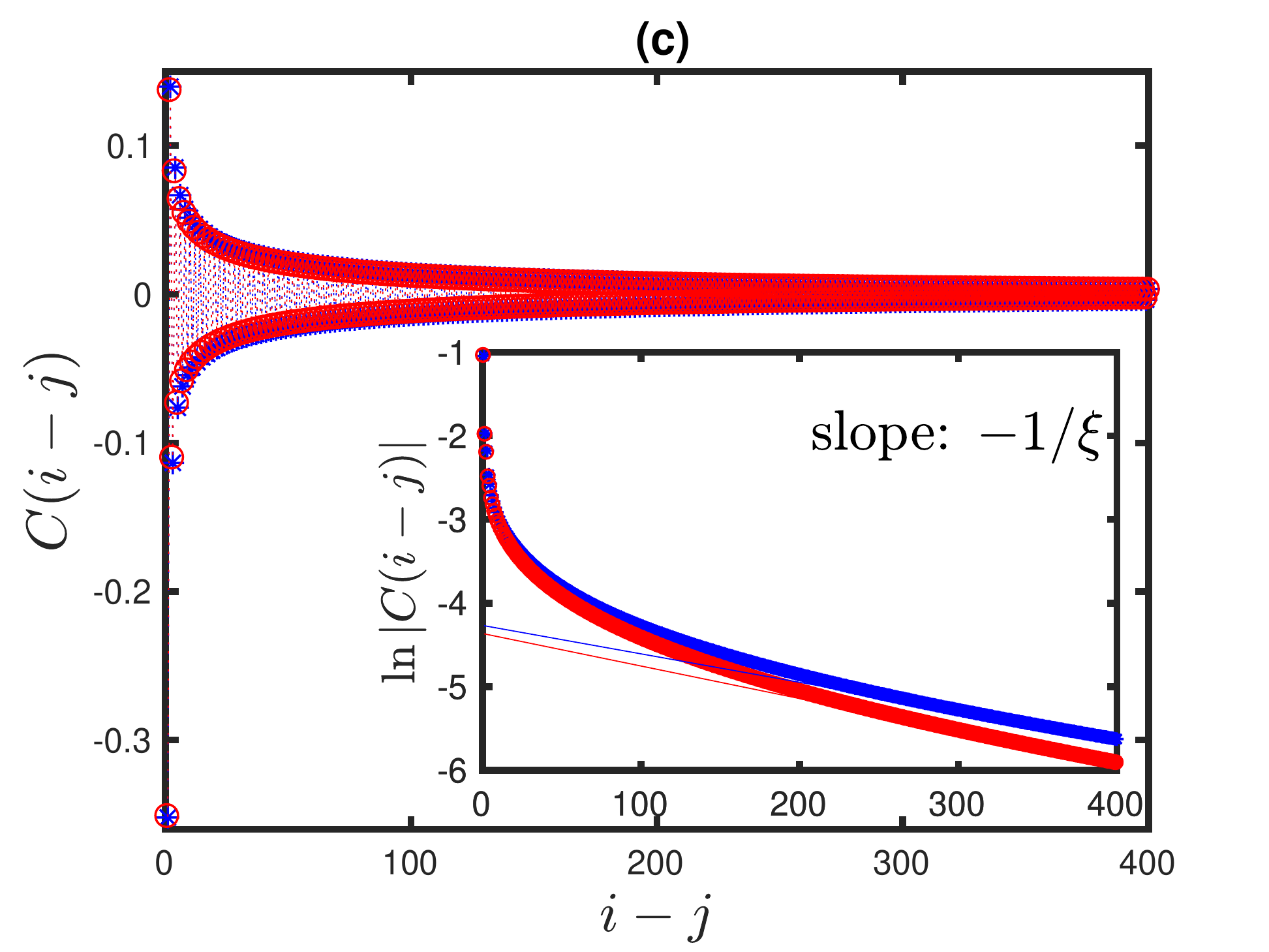}
\includegraphics[width=0.35\textwidth 
]{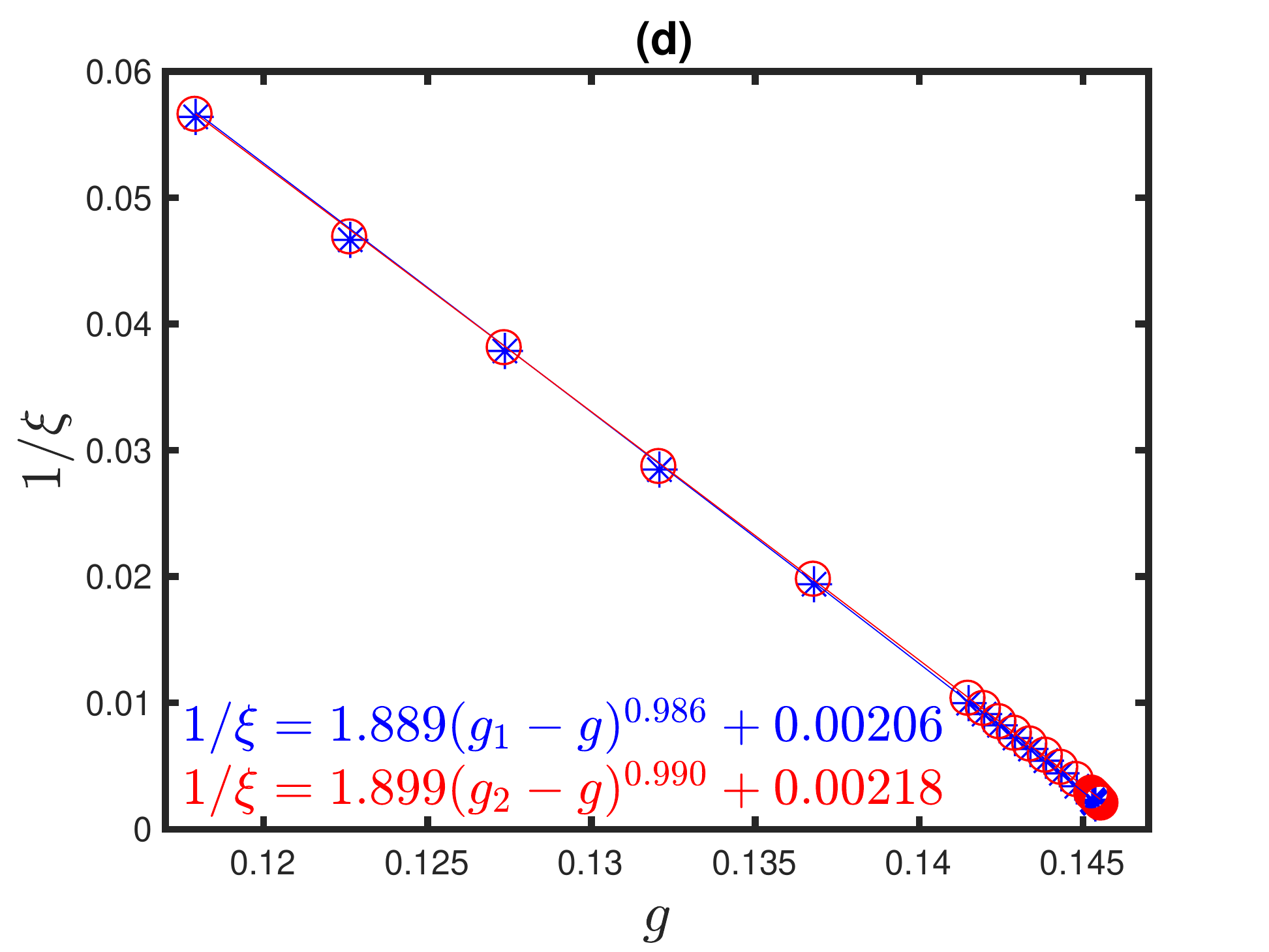}
\caption{Results from iTEBD calculation: (a) The average of
staggered magnetization $M_z$  vs. $g$ for cases with (blue) and without (red) the $h_z$ term.
The 
curves are fits to $M_z=0.524(g_{c1}-g)^{0.126}$ and $M_z=0.530(g_{c2}-g)^{0.128}$
for the two cases, respectively, where the critical points $g_{c1}=0.1454$ and $g_{c2}=0.1456$.
In either case 
$\beta\sim 1/8$ is obtained within error bar. The inset shows the
scaling plot of $M_z$ 
in the log-log scale. 
(b) Entanglement entropy $S_{\rm{EE}}(l)$ vs. chain 
segment length $l$ in the semi-log scale.
In either case the slope value agrees with a central charge of $c\sim1/2$.
(c) Spin-spin correlation function $C(i-j)$ vs. distance $i-j$ at $g=0.1448$ (near the 
QCP).
The inset indicates that $\ln C(i-j)$ is proportional to $i-j$ at large distance. (d) The inverse of correlation length $1/\xi$ vs. $g$,
where both cases give $\nu\sim1$.}
\label{fig:model1}
\end{figure*}

At small magnetic fields 
the ground state of the model is an 
antiferromagnetic (AF) ordered state. 
The applied magnetic field in $[100]$ direction plays the role of suppressing this order, and gives rise to a quantum phase transition at a QCP, $g_c$.
The induced staggered $h_y$ field enhances this effect, while the $h_z$ field has negligible contribution. To see this, we calculate the average magnetization in $z$ direction, $M_{z}$, using the iTEBD method for the Hamiltonian in Eq.~(\ref{eq:model1})
for two cases: with and without the $h_z$ term. We find that the $h_z$ term indeed only slightly changes the location of the QCP: 
As shown in Fig.~\ref{fig:model1}(a), the critical field $g_{c1}=0.1454(1)$ in the former case, and $g_{c2}=0.1456(1)$ for the latter. The critical exponent $\beta$ can be obtained by fitting $M_z\sim |g_{ci}-g|^\beta$ ($i=1,2$)
for both cases. In Fig.~\ref{fig:model1}(a) 
the fitting gives $\beta=0.126(5)$ and $0.128(5)$
for the two cases, both 
agree with $\beta\sim1/8$ of the 1DTFIM universality within error bars. We also calculate the entanglement entropy $S(l)=-\rm{Tr}\rho_A\ln\rho_A$ for a subsystem $A$ with $l$ spins in an infinite chain, where the reduced density matrix of $A$ is defined as $\rho_A=\rm{Tr_B}|\Psi_A\otimes\Psi_B\rangle\langle\Psi_A\otimes\Psi_B|$. (The ground state wavefunction can be decoupled to the tensor product of wavefunctions of subsystems, $\Psi_A$ and $\Psi_B$). From conformal field theory analysis~\cite{Cardy2004}, the entanglement entropy scales as $S(l)=\frac{c}{3}\ln l +\rm{const}$,
where $c$ is the central charge associated with the conformal invariance. Figure~\ref{fig:model1}(b) shows the scaling of $S(l)$ 
with subsystem size $l$,
which gives $c=0.502(5)$ and $c=0.503(5)$, for both cases with and without the $h_z$ term, respectively. Note that either 
agrees with $c=1/2$ of the 1DTFIM.
The critical exponent $\beta$ and central charge $c$
are consistent with Ref.~\cite{Giamarchi2018} where only the staggered field ($h_y$ term) is considered. We further calculate the spin-spin correlation,
denoted as $C(i-j)=\langle {S}_i^z {S}_j^z\rangle-\langle {S}_i^z \rangle\langle {S}_j^z \rangle$,
for Hamiltonian Eq.~(\ref{eq:model1}) with and without the $h_z$ term (see Fig.~\ref{fig:model1}(c)).
Using the relation $C(i-j)=e^{-|i-j|/\xi}$, with correlation
length $\xi$ and $1/\xi\sim |g_c-g|^{\nu}$, the critical exponent $\nu$ can be obtained. We find $\nu=0.986$ (0.990)$\sim1$ for the cases with (without) $h_z$ term (Fig.~\ref{fig:model1}d).

Based on these evidences, we conclude:
(1) The quantum phase transition falls in the 1DTFIM university class;
(2) 
the four-period $z$ field is an irrelevant term for this transition, which only slightly shifts the location of the QCP.
For the case when the transverse field is applied along the $[110]$ direction in 
some experimental setup, after some slight modifications, the model in Eq.~\eqref{eq:model1} can still describe the phase transition. In this case, we need to set $h_y=0$, and take the $h_z$ term 
to be $\cos(\frac{\pi (i-1)}{2})S_i^z$ for spin on site $i$,
with a four-period $\uparrow 0\downarrow 0$ pattern~\cite{KimuraJPSJ2013}.
Compare with the four-period $\uparrow\downarrow\downarrow\uparrow$ pattern for
the $[100]$ 
cases, the 
effects of the $\uparrow 0 \downarrow 0$ pattern of
the $h_z$ term is even weaker. 
In particular, the quantum phase transition still keeps the 1DTFIM university. 

\section{The generalized model}
\label{generalized}

Since the small $h_z$ term in Eq.~(\ref{eq:model1})
is irrelevant to 
the quantum phase transition induced by the transverse field, we drop the $h_z$ term and focus on the effects of the two inter-perpendicular transverse fields via studying a generalized model from the Hamiltonian in Eq.~\eqref{eq:model1}. 

\begin{eqnarray}
\label{eq:model2}
\nonumber
H_g&=&\sum_iJ[\epsilon(S^x_iS^x_{i+1}+S^y_iS^y_{i+1})+S^z_iS^z_{i+1}] \\
 &-&\sum_i [g_xS^x_i+g_y(-1)^iS^y_i]
\end{eqnarray}
with $\epsilon< 1$.  We consider arbitrary ratio $g_y/g_x$.
It extends the parameter region of the realistic model Eq.~(\ref{eq:model1})
in Sec.~\ref{model1},
where the ratio $h_y=g_y/g_x$ is fixed as 0.4. For small $g_x$ and $g_y$, the
Hamiltonian Eq.~(\ref{eq:model2}) has an AF ordered ground state.

\subsection{iTEBD calculation}
We first use the iTEBD method to study the model Eq.~(\ref{eq:model2}) extensively.
The average magnetization $M_z$ for 
various $g_x$ and $g_y$ values is calculated, and the quantum critical point is reached 
when the AF ordering is 
completely suppressed.

\begin{figure}
\begin{center}
\includegraphics[width=1\columnwidth]{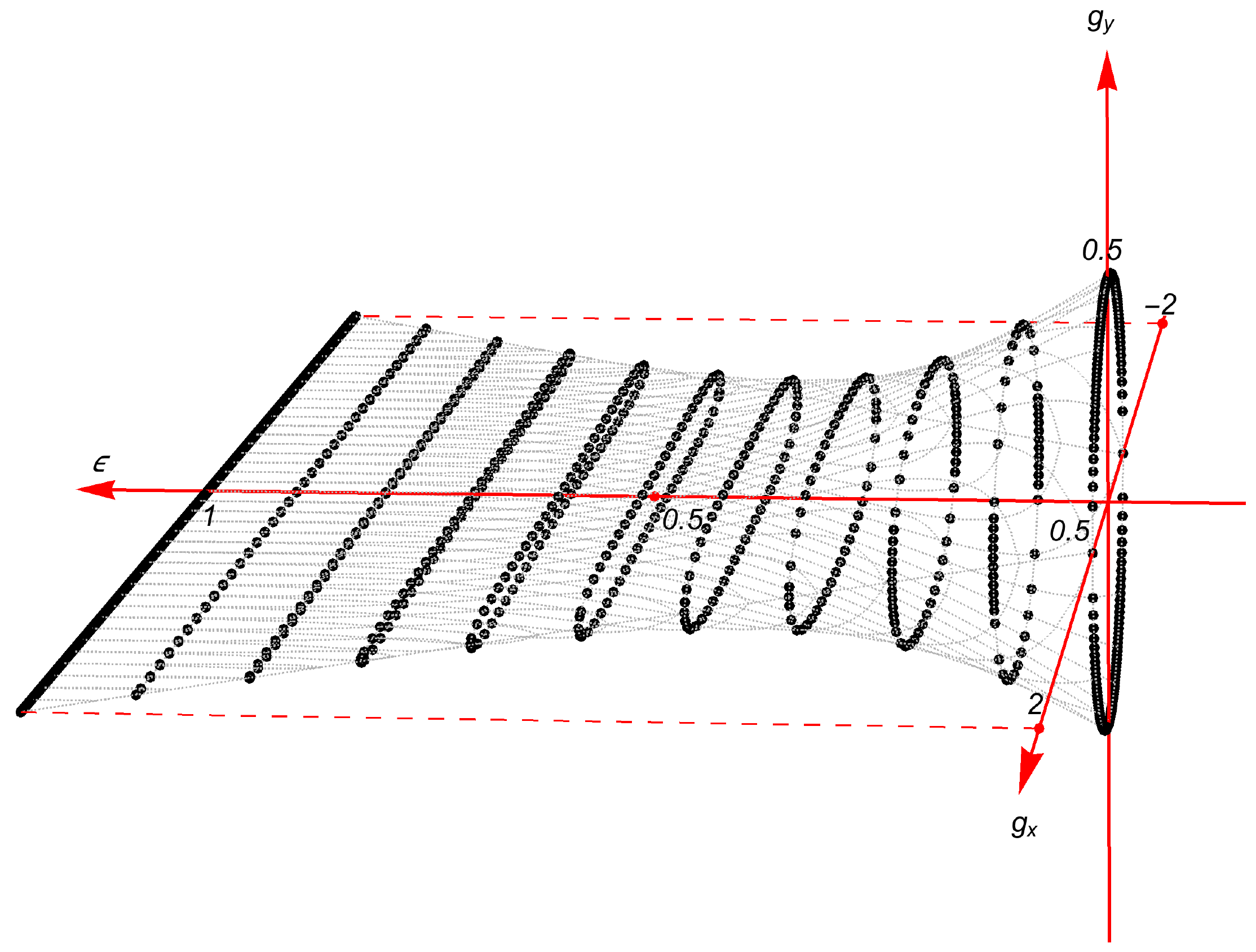}
\end{center}
\caption{The critical surface 
of the generalized model in Eq.~(\ref{eq:model2}) in the 3D parameter space from iTEBD calculation.}
\label{fig:critical_surface}
\end{figure}

We scan the parameters $g_x$ and $g_y$ with $\epsilon=$0,0.1,...,1, and obtain the QCPs of the generalized model in a 3D $g_x$-$g_y$-$\epsilon$ parameter space. As shown in Fig.~\ref{fig:critical_surface}, the obtained QCPs form a smooth critical surface. At $\epsilon=$0, \emph{i.e.} without the XY spin interaction terms,
the critical points in the $g_x$-$g_y$ plane are located on a circle with 
$|g_c|=0.5$ since in this case the model 
can be unitary transformed to the 1DTFIM. 
At a small $\epsilon$, the critical 
points approximately form an
ellipse with $g_{cx}>g_{cy}$, where $g_{c(x,y)}$ is the critical point on the $x$($y$) axis.
For 
$\epsilon\ge 0.4$, a nonmonotonic behavior of the critical
curve appears due to the nontrivial competition among the XY spin interactions,
$g_x$, and $g_y$ terms.
The critical curve forms a dumbbell shape with $g_{cx}\gg g_{cy}$.
$g_{cy}$ continuously shrinks with increasing $\epsilon$ and vanishes
at $\epsilon=1$ and $g_{cx}=2$. When $\epsilon=1$, the XXZ Heisenberg-Ising
Hamiltonian becomes the Heisenberg XXX model. Without the external fields,
the XXX model naturally contains an SU(2) symmetry with no long-range magnetic order
at zero temperature. Therefore, at $\epsilon = 1$ 
the external fields just simply
lead to paramagnetic response.
We notice that the 1DTFIM-type QCP obtained from Sec.~\ref{model1}
falls into this critical surface. By checking the
linearity between the magnetization $M^8_{\rm{z}}$ and $g_{x,y}$ across the QCPs
on the critical surface, we can conclude that
the 
transition across any point on the critical surface belongs to the 1DTFIM universality class.
The critical surface is symmetric along the $x$ (or $y$) axis as the
Hamiltonian is invariant by sending $S^{x(y)}\rightarrow -S^{x(y)}$ when $g_{x(y)}\rightarrow -g_{x(y)}$.
For $\epsilon>1$, the XXZ model will become Heisenberg-XY type
with physics dominated by the XY term. The influence of various external fields
at this situation will be deferred to future study.

\subsection{Mean field calculation}

To further understand the properties of the phase transition obtained from iTEBD calculation, a $\pi$ rotation of spins on even sites ($U=\exp[-i\pi\sum_jS^x_{2j}]$)
are carried out to transform the Hamiltonian [Eq.~(\ref{eq:model2})] into

\begin{eqnarray}
\label{eq:model3}
\nonumber
H_g'&=&\sum_iJ[\epsilon(S^x_iS^x_{i+1}-S^y_iS^y_{i+1})-S^z_iS^z_{i+1}] \\
 &-&\sum_i [g_xS^x_i+g_yS^y_i].
\end{eqnarray}
Correspondingly, the AF ground state at small
$g_x$ and $g_y$ values for Eq.~(\ref{eq:model2}) is transferred to a
Ferromagnetic (FM) state. Due to the asymmetric $\epsilon$ terms in Eq.~\eqref{eq:model3},
the perturbation of $g_x$ and $g_y$ is nonequivalent.
Both the $\epsilon$ and $g_{x,y}$ terms suppress the FM state,
but the $g_xS^x_i$ term competes with $\epsilon S^x_iS^x_{i+1}$.
It can induce a reentrant behavior of the staggered magnetization in the transverse XXZ model with $g_y=0$, which is discussed in Ref~\cite{YasuhiroPRB2001}
for $\epsilon\ge0.4$. However, the $g_yS^y_i$ 
and $-\epsilon S^y_iS^y_{i+1}$ terms have similar effects.
Thus, adding a $g_y$ term can efficiently suppress the FM phase.
This also means the staggered $y$-field can greatly assist to destroy
the AF ordering in the original model [Eq.~(\ref{eq:model2})].

We then carry out a mean field treatment~\cite{Dmitriev2002,CauxPRB2003}
to the generalized model [Eq.~(\ref{eq:model3})].
Here, another unitary transformation, rotating the spins around the $z$
axis by $\theta$, is performed, before rotating all the spins around the $x$ axis by $\pi/2$.
Through these two 
transformation, magnetic
field is directed along a new $z$ axis contributed
from both $g_x$ and $g_y$ terms. After a Jordan-Wigner transformation, a spinless fermion Hamiltonian is obtained,

\begin{eqnarray}
\nonumber
H''&=&\sum_i\frac{J_+}{2}(a^\dagger_ia_{i+1}+{\rm h.c.})+\frac{J_-}{2}(a^\dagger_ia^\dagger_{i+1}+\rm{h.c.})\\
&-&\gamma(n_i-1/2)(n_{i+1}-1/2)+g_0(n_i-1/2)
\end{eqnarray}
with $J_{\pm}=(\gamma\pm 1)/2$, $\gamma=\epsilon\cos(2\theta)$, $g_0=g_x\sin\theta+g_y\cos\theta$,
and $n_i=a^\dagger_ia_i$ is the density operator.
The four-fermion interaction term is decoupled in three possible ways, with the expectation values for the magnetization, kinetic hopping, and pairing terms denoted as
\begin{equation}
M=\langle n_i\rangle-1/2, K=\langle a^\dagger_{i+1}a_i\rangle, P=\langle a_{i+1}a_i\rangle.
\end{equation}

The mean-field Hamiltonian then becomes

\begin{equation}
H_{\rm{MF}}=\sum_i[\frac{\tilde{J}_+}{2}a^\dagger_ia_{i+1}+\frac{\tilde{J}_-}{2}a^\dagger_ia^\dagger_{i+1}+{\rm{h.c.}}+\tilde{g}(n_i-1/2)],
\end{equation}
where
\begin{equation}
\tilde{J}_+=J_++2\gamma K, \tilde{J}_-=J_--2\gamma P, \tilde{g}=g_0-2\gamma M.
\end{equation}
Fourier transform the Hamiltonian to the momentum space and diagonalize it using a Bogoliubov transformation,

 \begin{eqnarray}
 \nonumber
a_k&=&u_k\alpha_k+v_k\alpha^\dagger_{-k} \\
a^\dagger_{-k}&=&-v^*_k\alpha_k+u_k\alpha^\dagger_{-k}
 \end{eqnarray}
 with
 \begin{eqnarray*}
 u_k&=&\sqrt{(1+\frac{\tilde{J}_+\cos k+\tilde{g}}{\omega(k)})/2},\\
 v_k&=&i\;{\rm sgn}(k)\sqrt{(1-\frac{\tilde{J}_+\cos k+\tilde{g}}{\omega(k)})/2}.
 \end{eqnarray*}
The mean field Hamiltonian density becomes
\begin{eqnarray}
\nonumber
{H_{\rm{MF}}}/{N}&=&\int^\pi_0\frac{dk}{2\pi}[\frac{1}{2}\tilde{J}_+\cos k+\omega(k)(\alpha^\dagger_k\alpha_k-\frac{1}{2})] \\
&+&\gamma M^2-\gamma K^2+\gamma P^2,
\end{eqnarray}
where $\omega(k)$ is the single particle excitation spectrum,
\begin{equation}
\omega(k)=\sqrt{(\tilde{J}_+\cos k+\tilde{g})^2+(\tilde{J}_-)^2}.
\end{equation}

The quantities $M$,$K$,$P$ can be determined by the self-consistency conditions:
\begin{eqnarray*}
M&=&-\int^\pi_0\frac{dk}{2\pi}\frac{\tilde{J}_+\cos k+\tilde{g}}{\omega(k)},\\
K&=&-\int^\pi_0\frac{dk}{2\pi}\frac{(\tilde{J}_+\cos k+\tilde{g})\cos k}{\omega(k)},\\
P&=&-\int^\pi_0\frac{dk}{2\pi}\frac{\tilde{J}_-\sin^2 k}{\omega(k)},
\end{eqnarray*}
and the optimized $\theta$ can be fixed by minimizing the ground state energy.

\begin{figure}
\includegraphics[width=1\columnwidth]{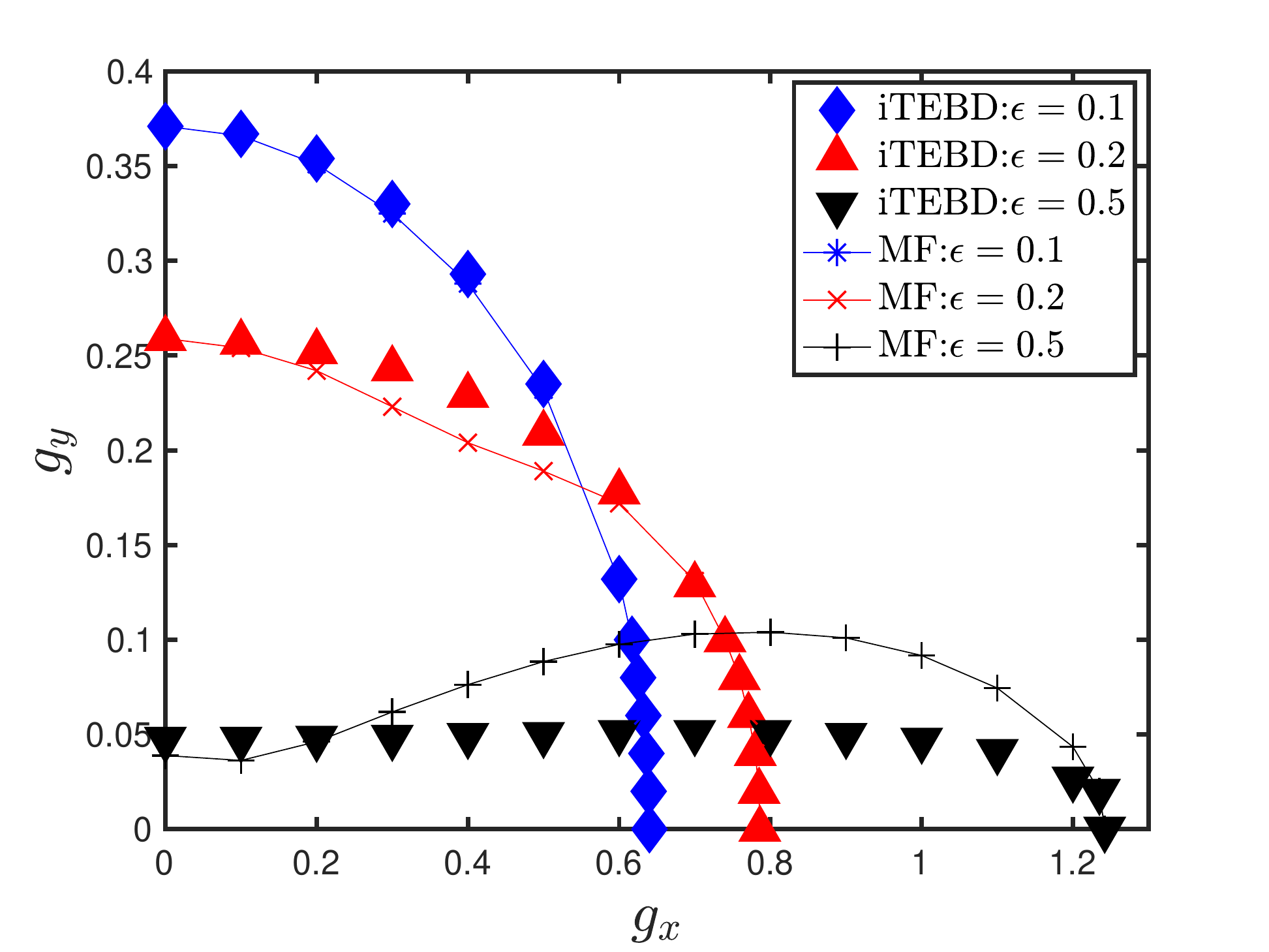}
\caption{
Critical line in the $g_x$-$g_y$ plane at $\epsilon=$0.1,0.2 and 0.5 from mean-field calculation (lines) and iTEBD (points).}
\label{fig:MF}
\end{figure}

The phase diagram in $g_x$-$g_y$ plane for a few different $\epsilon$ values
is shown in Fig.~\ref{fig:MF}. For small $\epsilon$, the critical line
is consistent with that determined from the iTEBD method. 
For $\epsilon=0.5$,
the nonmonotonic behavior of the transition line is also observed at mean-field level, although the critical line determined is quantitatively deviated from the iTEBD results. 

\begin{figure}
\includegraphics[width=0.49\columnwidth]{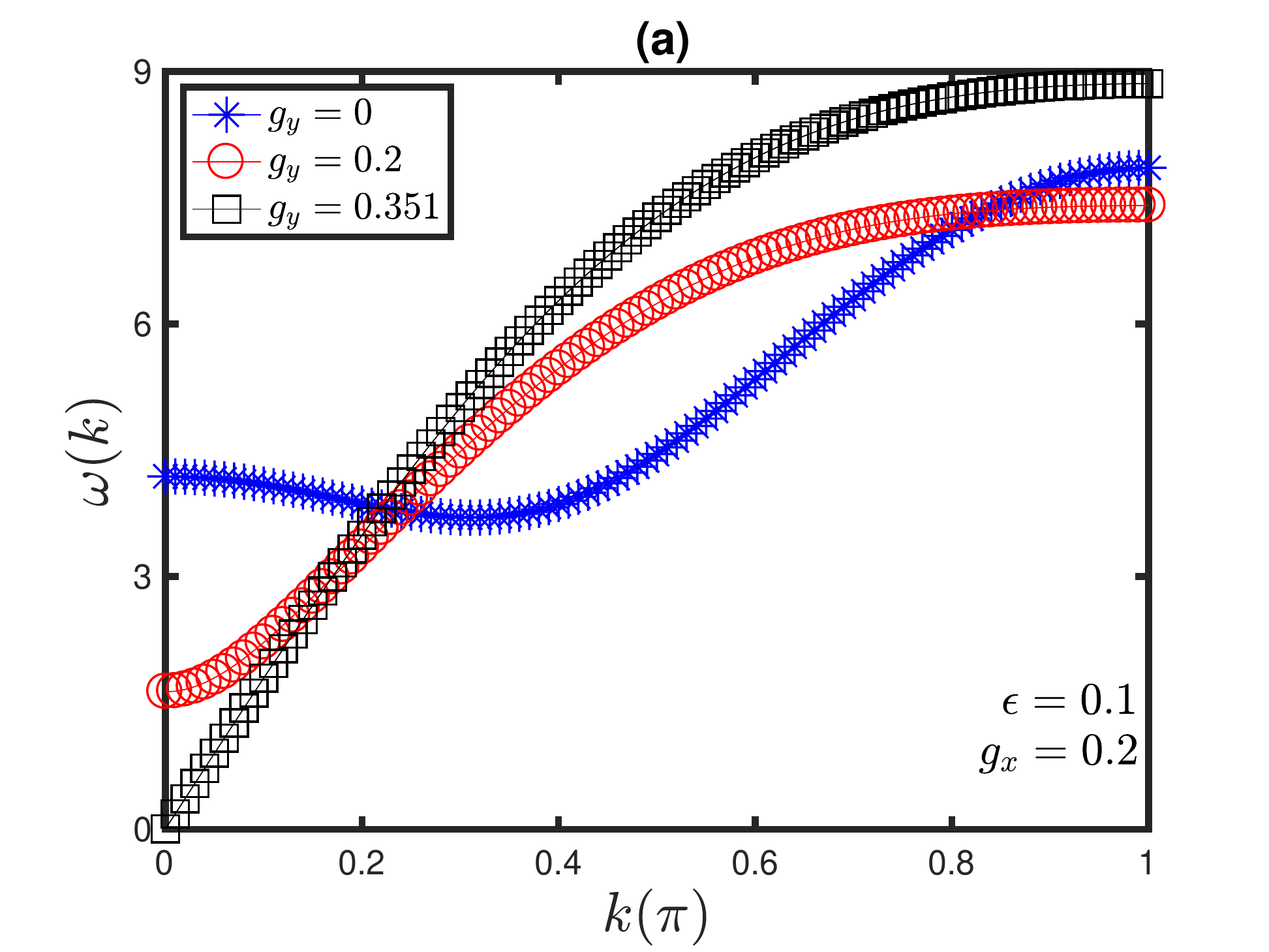}
\includegraphics[width=0.49\columnwidth]{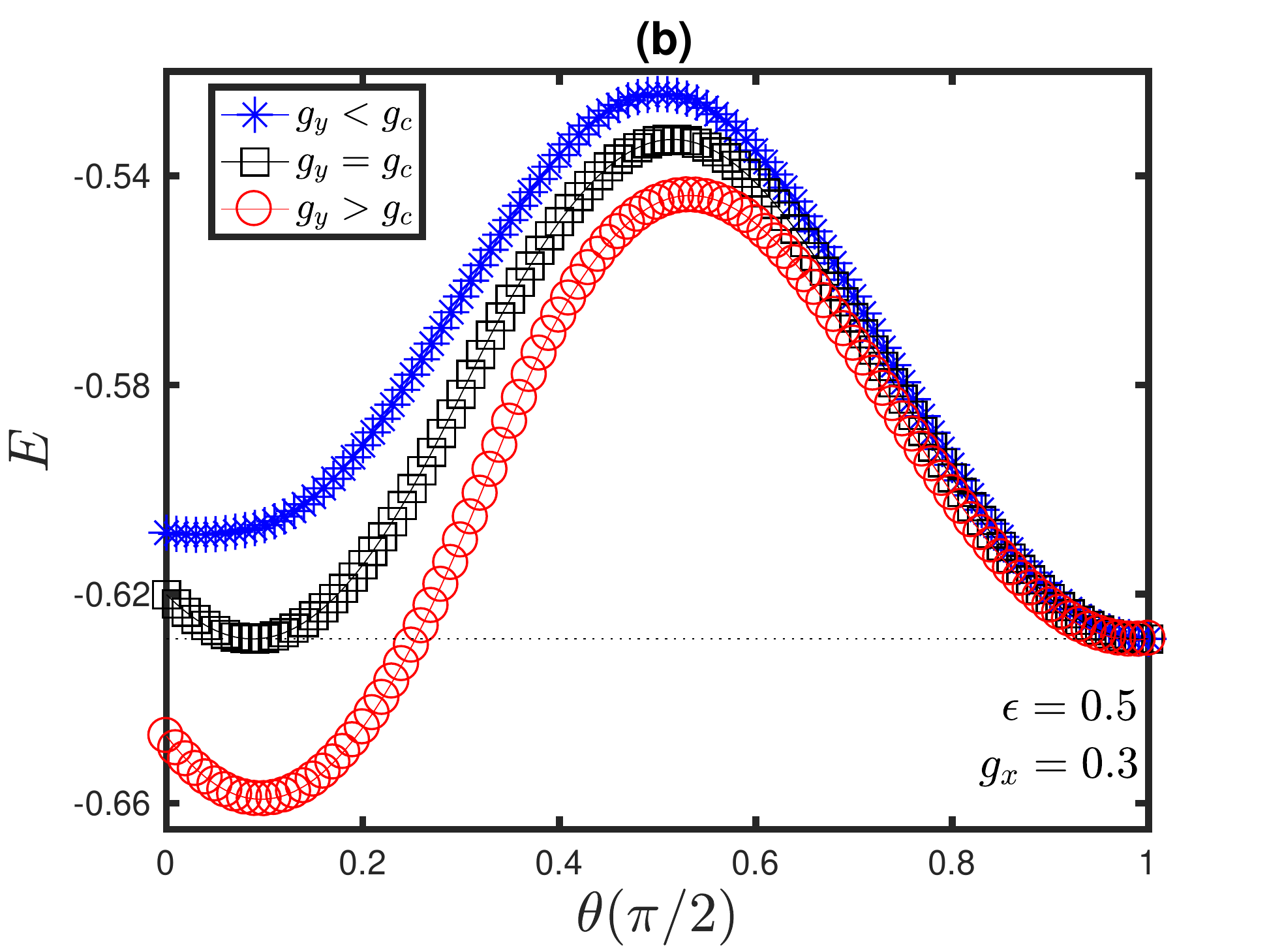}
\caption{(a) $\omega(k)$ as a function of $k$ at $\epsilon=0.1$ and $g_x=0.1$ for 
$g_y=0$,$0.2$, and $0.351$(transition point). (b) Double well structure of the energy as a function of $\theta$ at $\epsilon=0.5$ and $g_x=0.3$ for varied $g_y<g_c$, $g_y=g_c$, and $g_y>g_c$, where $g_c$ is the transition point}
\label{fig:MFE}
\end{figure}

We can understand the different critical line shapes at small and large $\epsilon$ values within the mean-field approach.
For small $\epsilon\le0.2$, there is only one optimized $\theta$ value
from 0 to $\pi/2$, while the parameter $g_x$ and $g_y$ are tuned from ($g_x=0,g_y>0$) to ($g_x>0,g_y=0$).
The single particle spectrum function $\omega(k)$ is gapless at $k=0$ at
the order-disorder transition point (Fig.~\ref{fig:MFE}(a)). For intermediate $\epsilon\ge0.5$, however, a double well structure of the energy as a function of $\theta$ appears (Fig.~\ref{fig:MFE}(b)), which
indicates competing quantum fluctuations 
between the $g_x$ and $g_y$ terms.
By fixing $g_x$, at $g_y<g_c$ where $g_c$ is the critical point,
the minimal energy point is closer to $\theta\sim\pi/2$, and thus
$g_0\sim g_x$. In this case, the $g_x$ term dominates the critical quantum
fluctuations. However, after $g_y$ crosses the critical point $g_c$,
the minimal energy point jumps to $\theta\sim 0$ and thus $g_0\sim g_y$,
which indicates that the $g_y$ term begins to play the major role.
We fix the phase transition boundary at the point $g_c$ where
the ground-state degenerate appears.
The above analysis based on the introduced optimized $\theta$
gives a qualitative explanation of the novel behavior of the
critical curve in the $g_x$-$g_y$ plane for large $\epsilon$.
At large $\epsilon$, the quantum fluctuations due to the XY spin interaction term is highly relevant,
while the interplay between the intrinsic fluctuations and the $g_{x,y}$
term makes the phase transition further nontrivial.
In the mean-field treatment, the odd-order terms $a_i^\dagger a_ia_{i+1}$ via
Jordan-Wigner transformation are neglected. The increasing discrepancy between
the mean field analysis and the iTEBD results with increasing $\epsilon$
indicates these neglected non-linear terms play a non-negligible role, and eventually
lead to the breakdown of the mean-field theory 
when $\epsilon \to 1$.

\section{Guidance to 
Experiments}
\label{experiment}
Because of the induced four-period $z$ field is irrelevant,
our extensive study of the phase transition on the generalized model
with broad parameter region $g_x$ and $g_y$ provides a concrete guidance
to the experiments on real materials of quasi-1D Ising anisotropic quantum magnets, such as 
SCVO and BCVO. 
In SCVO and BCVO, different $g_y/g_x$ ratio can be realized by applying a transverse
field which gradually rotates in the $x-y$ plane. The rotation of the applied transverse field
shifts the position of the QCP. At the $[100]$ direction,
strongest in-plane staggered field $g_y$ perpendicular to the
external transverse one is induced. Deviation of the applied transverse
field from the $[100]$ direction weakens $g_y$.
Consequently it requires stronger applied transverse field to reach the 1DTFIM QCP.
And when the applied field reaches the direction of $[110]$,
the required external transverse field reaches maximum.
This is clearly observed in BCVO~\cite{WangPRL2018} and the recent
NMR experiments on SCVO~\cite{YuRecent} with the external
transverse field applied along [110] and [100], respectively.
For SCVO, when the field is applied to the
$[100]$ direction
the 1D QCP of SCVO is slightly outside the 3D ordered phase~\cite{YuRecent}.
As the QCP(1D) will be pushed further by rotating the in-plane applied field,
it can be used to 
distinguish unambiguously the 3D QCP and 1D QCP at
$[100]$ direction and provide a clean platform to examine
the novel quantum critical behavior near the 1D QCP.

\section{Conclusion}
\label{conclusion}
To conclude, following the corresponding effective
one-dimensional Heisenberg-Ising model with axial anisotropy
tuned by $\epsilon$,
we demonstrate that the observed transverse-field ($h_x$)-tuned
quantum phase transition outside the three-dimensional N\'{e}el order
in the materials of BCVO and SCVO falls into the 1DTFIM universality, despite
of the presence of 
the induced in-plane staggered field ($h_y$) and
the 
four-period fields along the $c$ axis ($h_z$).
Small $h_z$ is further shown to make negligible
effects on the phase transition. A generalized model is immediately constructed
by neglecting $h_z$ and relaxing the
$\epsilon$, $h_x$, and $h_y$
to a broad parameter region. 
With the generalized model,
the nature of quantum phase transitions is then extensively and carefully
scrutinized by the iTEBD method, and 
it has been shown that the quantum phase transition falls into the 1DTFIM universality,
rendering out a nice 1DTFIM quantum critical surface
until $\epsilon = 1$ (XXX limit).
The well-established 1DTFIM critical surface obtained
from the generalized model is expected to guide
the in-plane-field-rotation measurements
on real materials such as SCVO, BCVO, and CoNb$_2$O$_6$ {\it etc}..
Our study concretely
demonstrates that the 1DTFIM universality 
is robust against additional spin interaction terms as well as various external
fields. This opens rich opportunities 
to access to the 1DTFIM QCP and realize the celebrated
Zamolodchikov quantum $E_8$ integrable
model~\cite{E8_1989} near this QCP in real materials.

\section{Acknowledgments}
We thank Weiqiang Yu and Zhe Wang for helpful discussions.
The work at Shanghai Jiao Tong University is supported by the National Natural Science
Foundation of China Grant No. 11804221 (H.Z.), and Science
and Technology Commission of Shanghai Municipality
Grant No. 16DZ2260200 (H.Z.).
The work at Renmin University of China was supported by the Ministry of Science and Technology of China, National Program on Key Research Project Grant number 2016YFA0300504,
the National Science Foundation of China Grant number 11674392, and the Fundamental Research Funds for the Central Universities and the Research Funds of Remnin University
of China Grant number 18XNLG24 (R.Y.). 
J.W. acknowledges support from Shanghai city.

\appendix*
\section{iTEBD results}
In this appendix, we list the values of the quantum critical points,
used to generate the 
critical surface in Fig.~2 of the main text.
In tables I-III, $[g_x,g_y]$ are the critical points in the $g_x$-$g_y$ plane
for different $\epsilon$. All the critical points are obtained
by scanning the parameter space along $g_x$ or $g_y$
with one of them held at fixed value. The width of error bars is
the step size of the iTEBD calculation.

\begin{table}[th]
\begin{tabular}{|c|c|c|}
\hline
\hline
$\epsilon=0.1$&$\epsilon=0.2$&$\epsilon=0.3$\cr
\hline
$[0, 0.371(1)]$&$[0, 0.259(1)]$&$[0, 0.166(1)]$\cr
\hline
$[0.1, 0.367(1)]$&$[0.1, 0.257(1)]$&$[0.1, 0.166(1)]$\cr
\hline
$[0.2, 0.354(1)]$&$[0.2, 0.252(1)]$&$[0.2, 0.164(1)]$\cr
\hline
$[0.3, 0.330(1)]$&$[0.3, 0.243(1)]$&$[0.3, 0.161(1)]$\cr
\hline
$[0.4, 0.293(1)]$&$[0.4, 0.229(1)]$&$[0.4, 0.157(1)]$\cr
\hline
$[0.5, 0.235(1)]$&$[0.5, 0.209(1)]$&$[0.5, 0.151(1)]$\cr
\hline
$[0.6, 0.132(1)]$&$[0.6, 0.178(1)]$&$[0.6, 0.141(1)]$\cr
\hline
$[0.617(1), 0.1]$&$[0.7, 0.129(1)]$&$[0.7, 0.127(1)]$\cr
\hline
$[0.625(1), 0.08]$&$[0.736(1), 0.1]$&$[0.8, 0.104(1)]$\cr
\hline
$[0.631(1), 0.06]$&$[0.755(1), 0.08]$&$[0.9, 0.058(1)]$\cr
\hline
$[0.636(1), 0.04]$&$[0.769(1), 0.06]$&$[0.932(1), 0.02]$\cr
\hline
$[0.639(1), 0.02]$&$[0.778(1), 0.04]$&$[0.936(1), 0]$\cr
\hline
$[0.639(1), 0]$&$[0.784(1), 0.02]$&$----$\cr
\hline
$----$&$[0.786(1), 0]$&$----$\cr
\hline
\hline
\end{tabular}
\caption{\label{tb:pp1} Phase transition points in $g_x$-$g_y$ plane from iTEBD calculation.}
\end{table}

\begin{table}[th]
\begin{tabular}{|c|c|c|}
\hline
\hline
$\epsilon=0.4$&$\epsilon=0.5$&$\epsilon=0.6$\cr
\hline
$[0, 0.095(1)]$&$[0, 0.0475(5)]$&$[0, 0.0190(5)]$\cr
\hline
$[0.1, 0.095(1)]$&$[0.1, 0.0475(5)]$&$[0.1, 0.0190(5)]$\cr
\hline
$[0.2, 0.095(1)]$&$[0.2, 0.0480(5)]$&$[0.2, 0.0195(5)]$\cr
\hline
$[0.3, 0.095(1)]$&$[0.3, 0.0485(5)]$&$[0.3, 0.0200(5)]$\cr
\hline
$[0.4, 0.095(1)]$&$[0.4, 0.0495(5)]$&$[0.4, 0.0210(5)]$\cr
\hline
$[0.5, 0.094(1)]$&$[0.5, 0.0500(5)]$&$[0.5, 0.0220(5)]$\cr
\hline
$[0.6, 0.092(1)]$&$[0.6, 0.0510(5)]$&$[0.6, 0.0230(5)]$\cr
\hline
$[0.7, 0.089(1)]$&$[0.7, 0.0510(5)]$&$[0.7, 0.0240(5)]$\cr
\hline
$[0.8, 0.083(1)]$&$[0.8, 0.0510(5)]$&$[0.8, 0.0250(5)]$\cr
\hline
$[0.9, 0.074(1)]$&$[0.9, 0.0495(5)]$&$[0.9, 0.0255(5)]$\cr
\hline
$[1.0, 0.057(1)]$&$[1.0, 0.0470(5)]$&$[1.0, 0.0260(5)]$\cr
\hline
$[1.050(1), 0.04]$&$[1.1, 0.0410(5)]$&$[1.1, 0.0260(5)]$\cr
\hline
$[1.080(1), 0.02]$&$[1.2, 0.0265(5)]$&$[1.2, 0.0245(5)]$\cr
\hline
$[1.089(1), 0]$&$[1.220(1), 0.02]$&$[1.3, 0.0205(5)]$\cr
\hline
$----$&$[1.242(1), 0]$&$[1.380(1), 0.01]$\cr
\hline
$----$&$----$&$[1.394(1), 0]$\cr
\hline
\hline
\end{tabular}
\caption{\label{tb:pp2} Phase transition points in $g_x$-$g_y$ plane from iTEBD calculation (continue).}
\end{table}

\begin{table}[th]
\begin{tabular}{|c|c|c|}
\hline
\hline
$\epsilon=0.7$&$\epsilon=0.8$&$\epsilon=0.9$\cr
\hline
$[0, 0.0056(2)]$&$[0, 0.0016(2)]$&$[0, 0.0004(2)]$\cr
\hline
$[0.1, 0.0058(2)]$&$[0.1, 0.0016(2)]$&$[0.1, 0.0004(2)]$\cr
\hline
$[0.2, 0.0060(2)]$&$[0.2, 0.0018(2)]$&$[0.2, 0.0004(2)]$\cr
\hline
$[0.3, 0.0064(2)]$&$[0.3, 0.0020(2)]$&$[0.3, 0.0004(2)]$\cr
\hline
$[0.4, 0.0070(2)]$&$[0.4, 0.0020(2)]$&$[0.4, 0.0004(2)]$\cr
\hline
$[0.5, 0.0074(2)]$&$[0.5, 0.0022(2)]$&$[0.5, 0.0004(2)]$\cr
\hline
$[0.6, 0.0080(2)]$&$[0.6, 0.0022(2)]$&$[0.6, 0.0004(2)]$\cr
\hline
$[0.7, 0.0086(2)]$&$[0.7, 0.0024(2)]$&$[0.7, 0.0004(2)]$\cr
\hline
$[0.8, 0.0094(2)]$&$[0.8, 0.0026(2)]$&$[0.8, 0.0006(2)]$\cr
\hline
$[0.9, 0.0100(2)]$&$[0.9, 0.0030(2)]$&$[0.9, 0.0006(2)]$\cr
\hline
$[1.0, 0.0108(2)]$&$[1.0, 0.0032(2)]$&$[1.0, 0.0006(2)]$\cr
\hline
$[1.1, 0.0114(2)]$&$[1.1, 0.0036(2)]$&$[1.1, 0.0006(2)]$\cr
\hline
$[1.2, 0.0120(2)]$&$[1.2, 0.0038(2)]$&$[1.2, 0.0008(2)]$\cr
\hline
$[1.3, 0.0120(2)]$&$[1.3, 0.0042(2)]$&$[1.3, 0.0008(2)]$\cr
\hline
$[1.4, 0.0116(2)]$&$[1.4, 0.0044(2)]$&$[1.4, 0.0008(2)]$\cr
\hline
$[1.5, 0.0086(2)]$&$[1.5, 0.0046(2)]$&$[1.5, 0.0010(2)]$\cr
\hline
$[1.536(1), 0.005]$&$[1.6, 0.0042(2)]$&$[1.6, 0.0010(2)]$\cr
\hline
$[1.547(1), 0]$&$[1.690(1), 0.002]$&$[1.7, 0.0010(2)]$\cr
\hline
$----$&$[1.698(1), 0]$&$[1.8, 0.0008(2)]$\cr
\hline
$----$&$----$&$[1.849(1), 0]$\cr
\hline
\hline
\end{tabular}
\caption{\label{tb:pp3} Phase transition points in $g_x$-$g_y$ plane from iTEBD calculation.}
\end{table}

Technically, convergent iTEBD results upto particular significant digit can be reached as the Schmidt rank $\chi$ which characterizes the entanglement of the system is increased. In Sec. II, we carry the iTEBD calculation with $\chi=10,20,..., 70$ to obtain convergent four significant digit results for QCP locations. However, To obtain convergent three significant digit results, $\chi=20$ is enough. Thus, the numerious QCP results in Sec. III is calculated from $\chi=20$.

\bibliography{1DIsing}

\begin{thebibliography}{43}%
\makeatletter
\providecommand \@ifxundefined [1]{%
 \@ifx{#1\undefined}
}%
\providecommand \@ifnum [1]{%
 \ifnum #1\expandafter \@firstoftwo
 \else \expandafter \@secondoftwo
 \fi
}%
\providecommand \@ifx [1]{%
 \ifx #1\expandafter \@firstoftwo
 \else \expandafter \@secondoftwo
 \fi
}%
\providecommand \natexlab [1]{#1}%
\providecommand \enquote  [1]{``#1''}%
\providecommand \bibnamefont  [1]{#1}%
\providecommand \bibfnamefont [1]{#1}%
\providecommand \citenamefont [1]{#1}%
\providecommand \href@noop [0]{\@secondoftwo}%
\providecommand \href [0]{\begingroup \@sanitize@url \@href}%
\providecommand \@href[1]{\@@startlink{#1}\@@href}%
\providecommand \@@href[1]{\endgroup#1\@@endlink}%
\providecommand \@sanitize@url [0]{\catcode `\\12\catcode `\$12\catcode
  `\&12\catcode `\#12\catcode `\^12\catcode `\_12\catcode `\%12\relax}%
\providecommand \@@startlink[1]{}%
\providecommand \@@endlink[0]{}%
\providecommand \url  [0]{\begingroup\@sanitize@url \@url }%
\providecommand \@url [1]{\endgroup\@href {#1}{\urlprefix }}%
\providecommand \urlprefix  [0]{URL }%
\providecommand \Eprint [0]{\href }%
\providecommand \doibase [0]{http://dx.doi.org/}%
\providecommand \selectlanguage [0]{\@gobble}%
\providecommand \bibinfo  [0]{\@secondoftwo}%
\providecommand \bibfield  [0]{\@secondoftwo}%
\providecommand \translation [1]{[#1]}%
\providecommand \BibitemOpen [0]{}%
\providecommand \bibitemStop [0]{}%
\providecommand \bibitemNoStop [0]{.\EOS\space}%
\providecommand \EOS [0]{\spacefactor3000\relax}%
\providecommand \BibitemShut  [1]{\csname bibitem#1\endcsname}%
\let\auto@bib@innerbib\@empty
\bibitem [{Spe(2010)}]{SpecialIssue2010}%
  \BibitemOpen
  \href {\doibase 10.1007/s10909-010-0219-y} {\bibfield  {journal} {\bibinfo
  {journal} {Special issue on Quantum Phase Transitions, J. Low Temp. Phys.}\
  }\textbf {\bibinfo {volume} {161}},\ \bibinfo {pages} {1} (\bibinfo {year}
  {2010})}\BibitemShut {NoStop}%
\bibitem [{\citenamefont {Sachdev}(2011)}]{sachdev2011}%
  \BibitemOpen
  \bibfield  {author} {\bibinfo {author} {\bibfnamefont {S.}~\bibnamefont
  {Sachdev}},\ }\href@noop {} {\emph {\bibinfo {title} {Quantum phase
  transitions}}},\ \bibinfo {edition} {second ed.}\ ed.\ (\bibinfo  {publisher}
  {Cambridge University Press},\ \bibinfo {address} {Cambridge},\ \bibinfo
  {year} {2011})\BibitemShut {NoStop}%
\bibitem [{\citenamefont {Si}\ \emph {et~al.}(2001)\citenamefont {Si},
  \citenamefont {Rabello}, \citenamefont {Ingersent},\ and\ \citenamefont
  {Smith}}]{Si2001}%
  \BibitemOpen
  \bibfield  {author} {\bibinfo {author} {\bibfnamefont {Q.}~\bibnamefont
  {Si}}, \bibinfo {author} {\bibfnamefont {S.}~\bibnamefont {Rabello}},
  \bibinfo {author} {\bibfnamefont {K.}~\bibnamefont {Ingersent}}, \ and\
  \bibinfo {author} {\bibfnamefont {J.~L.}\ \bibnamefont {Smith}},\ }\href@noop
  {} {\bibfield  {journal} {\bibinfo  {journal} {Nature}\ }\textbf {\bibinfo
  {volume} {413}},\ \bibinfo {pages} {804} (\bibinfo {year}
  {2001})}\BibitemShut {NoStop}%
\bibitem [{\citenamefont {Schr{\"o}der}\ \emph {et~al.}(2000)\citenamefont
  {Schr{\"o}der}, \citenamefont {Aeppli}, \citenamefont {Coldea}, \citenamefont
  {Adams}, \citenamefont {Stockert}, \citenamefont {L{\"o}hneysen},
  \citenamefont {Bucher}, \citenamefont {Ramazashvili},\ and\ \citenamefont
  {Coleman}}]{Schroder2000}%
  \BibitemOpen
  \bibfield  {author} {\bibinfo {author} {\bibfnamefont {A.}~\bibnamefont
  {Schr{\"o}der}}, \bibinfo {author} {\bibfnamefont {G.}~\bibnamefont
  {Aeppli}}, \bibinfo {author} {\bibfnamefont {R.}~\bibnamefont {Coldea}},
  \bibinfo {author} {\bibfnamefont {M.}~\bibnamefont {Adams}}, \bibinfo
  {author} {\bibfnamefont {O.}~\bibnamefont {Stockert}}, \bibinfo {author}
  {\bibfnamefont {H.}~\bibnamefont {L{\"o}hneysen}}, \bibinfo {author}
  {\bibfnamefont {E.}~\bibnamefont {Bucher}}, \bibinfo {author} {\bibfnamefont
  {R.}~\bibnamefont {Ramazashvili}}, \ and\ \bibinfo {author} {\bibfnamefont
  {P.}~\bibnamefont {Coleman}},\ }\href@noop {} {\bibfield  {journal} {\bibinfo
   {journal} {Nature}\ }\textbf {\bibinfo {volume} {407}},\ \bibinfo {pages}
  {351} (\bibinfo {year} {2000})}\BibitemShut {NoStop}%
\bibitem [{\citenamefont {Coleman}\ and\ \citenamefont
  {Schofield}(2005)}]{Coleman2005}%
  \BibitemOpen
  \bibfield  {author} {\bibinfo {author} {\bibfnamefont {P.}~\bibnamefont
  {Coleman}}\ and\ \bibinfo {author} {\bibfnamefont {A.~J.}\ \bibnamefont
  {Schofield}},\ }\href@noop {} {\bibfield  {journal} {\bibinfo  {journal}
  {Nature}\ }\textbf {\bibinfo {volume} {433}},\ \bibinfo {pages} {226}
  (\bibinfo {year} {2005})}\BibitemShut {NoStop}%
\bibitem [{\citenamefont {Si}\ and\ \citenamefont
  {Steglich}(2010)}]{Si_Science2010}%
  \BibitemOpen
  \bibfield  {author} {\bibinfo {author} {\bibfnamefont {Q.}~\bibnamefont
  {Si}}\ and\ \bibinfo {author} {\bibfnamefont {F.}~\bibnamefont {Steglich}},\
  }\href@noop {} {\bibfield  {journal} {\bibinfo  {journal} {Science}\ }\textbf
  {\bibinfo {volume} {329}},\ \bibinfo {pages} {1161} (\bibinfo {year}
  {2010})}\BibitemShut {NoStop}%
\bibitem [{\citenamefont {Schuberth}\ \emph {et~al.}(2016)\citenamefont
  {Schuberth}, \citenamefont {Tippmann}, \citenamefont {Steinke}, \citenamefont
  {Lausberg}, \citenamefont {Steppke}, \citenamefont {Brando}, \citenamefont
  {Krellner}, \citenamefont {Geibel}, \citenamefont {Yu}, \citenamefont {Si}
  \emph {et~al.}}]{Schuberth2016}%
  \BibitemOpen
  \bibfield  {author} {\bibinfo {author} {\bibfnamefont {E.}~\bibnamefont
  {Schuberth}}, \bibinfo {author} {\bibfnamefont {M.}~\bibnamefont {Tippmann}},
  \bibinfo {author} {\bibfnamefont {L.}~\bibnamefont {Steinke}}, \bibinfo
  {author} {\bibfnamefont {S.}~\bibnamefont {Lausberg}}, \bibinfo {author}
  {\bibfnamefont {A.}~\bibnamefont {Steppke}}, \bibinfo {author} {\bibfnamefont
  {M.}~\bibnamefont {Brando}}, \bibinfo {author} {\bibfnamefont
  {C.}~\bibnamefont {Krellner}}, \bibinfo {author} {\bibfnamefont
  {C.}~\bibnamefont {Geibel}}, \bibinfo {author} {\bibfnamefont
  {R.}~\bibnamefont {Yu}}, \bibinfo {author} {\bibfnamefont {Q.}~\bibnamefont
  {Si}},  \emph {et~al.},\ }\href@noop {} {\bibfield  {journal} {\bibinfo
  {journal} {Science}\ }\textbf {\bibinfo {volume} {351}},\ \bibinfo {pages}
  {485} (\bibinfo {year} {2016})}\BibitemShut {NoStop}%
\bibitem [{\citenamefont {Wu}\ \emph {et~al.}(2016)\citenamefont {Wu},
  \citenamefont {Si},\ and\ \citenamefont {Abrahams}}]{Wu2016b}%
  \BibitemOpen
  \bibfield  {author} {\bibinfo {author} {\bibfnamefont {J.}~\bibnamefont
  {Wu}}, \bibinfo {author} {\bibfnamefont {Q.}~\bibnamefont {Si}}, \ and\
  \bibinfo {author} {\bibfnamefont {E.}~\bibnamefont {Abrahams}},\ }\href@noop
  {} {\bibfield  {journal} {\bibinfo  {journal} {Physical Review B}\ }\textbf
  {\bibinfo {volume} {93}},\ \bibinfo {pages} {104515} (\bibinfo {year}
  {2016})}\BibitemShut {NoStop}%
\bibitem [{\citenamefont {Kinross}\ \emph {et~al.}(2014)\citenamefont
  {Kinross}, \citenamefont {Fu}, \citenamefont {Munsie}, \citenamefont
  {Dabkowska}, \citenamefont {Luke}, \citenamefont {Sachdev},\ and\
  \citenamefont {Imai}}]{Kinross2014}%
  \BibitemOpen
  \bibfield  {author} {\bibinfo {author} {\bibfnamefont {A.~W.}\ \bibnamefont
  {Kinross}}, \bibinfo {author} {\bibfnamefont {M.}~\bibnamefont {Fu}},
  \bibinfo {author} {\bibfnamefont {T.~J.}\ \bibnamefont {Munsie}}, \bibinfo
  {author} {\bibfnamefont {H.~A.}\ \bibnamefont {Dabkowska}}, \bibinfo {author}
  {\bibfnamefont {G.~M.}\ \bibnamefont {Luke}}, \bibinfo {author}
  {\bibfnamefont {S.}~\bibnamefont {Sachdev}}, \ and\ \bibinfo {author}
  {\bibfnamefont {T.}~\bibnamefont {Imai}},\ }\href {\doibase
  10.1103/PhysRevX.4.031008} {\bibfield  {journal} {\bibinfo  {journal} {Phys.
  Rev. X}\ }\textbf {\bibinfo {volume} {4}},\ \bibinfo {pages} {031008}
  (\bibinfo {year} {2014})}\BibitemShut {NoStop}%
\bibitem [{\citenamefont {Wu}\ \emph {et~al.}(2014)\citenamefont {Wu},
  \citenamefont {Kormos},\ and\ \citenamefont {Si}}]{WuPRL2014}%
  \BibitemOpen
  \bibfield  {author} {\bibinfo {author} {\bibfnamefont {J.}~\bibnamefont
  {Wu}}, \bibinfo {author} {\bibfnamefont {M.}~\bibnamefont {Kormos}}, \ and\
  \bibinfo {author} {\bibfnamefont {Q.}~\bibnamefont {Si}},\ }\href {\doibase
  10.1103/PhysRevLett.113.247201} {\bibfield  {journal} {\bibinfo  {journal}
  {Phys. Rev. Lett.}\ }\textbf {\bibinfo {volume} {113}},\ \bibinfo {pages}
  {247201} (\bibinfo {year} {2014})}\BibitemShut {NoStop}%
\bibitem [{\citenamefont {Niemeijer}(1967)}]{Niemeijer1967}%
  \BibitemOpen
  \bibfield  {author} {\bibinfo {author} {\bibfnamefont {T.}~\bibnamefont
  {Niemeijer}},\ }\href@noop {} {\bibfield  {journal} {\bibinfo  {journal}
  {Physica}\ }\textbf {\bibinfo {volume} {36}},\ \bibinfo {pages} {377}
  (\bibinfo {year} {1967})}\BibitemShut {NoStop}%
\bibitem [{\citenamefont {Pfeuty}(1970)}]{Pfeuty1970}%
  \BibitemOpen
  \bibfield  {author} {\bibinfo {author} {\bibfnamefont {P.}~\bibnamefont
  {Pfeuty}},\ }\href {\doibase https://doi.org/10.1016/0003-4916(70)90270-8}
  {\bibfield  {journal} {\bibinfo  {journal} {Annals of Physics}\ }\textbf
  {\bibinfo {volume} {57}},\ \bibinfo {pages} {79 } (\bibinfo {year}
  {1970})}\BibitemShut {NoStop}%
\bibitem [{\citenamefont {Barouch}\ and\ \citenamefont
  {McCoy}(1971)}]{McCoy1971}%
  \BibitemOpen
  \bibfield  {author} {\bibinfo {author} {\bibfnamefont {E.}~\bibnamefont
  {Barouch}}\ and\ \bibinfo {author} {\bibfnamefont {B.~M.}\ \bibnamefont
  {McCoy}},\ }\href@noop {} {\bibfield  {journal} {\bibinfo  {journal} {Phys.
  Rev. A}\ }\textbf {\bibinfo {volume} {3}},\ \bibinfo {pages} {786} (\bibinfo
  {year} {1971})}\BibitemShut {NoStop}%
\bibitem [{\citenamefont {Suzuki}(1971)}]{Suzuki1971}%
  \BibitemOpen
  \bibfield  {author} {\bibinfo {author} {\bibfnamefont {M.}~\bibnamefont
  {Suzuki}},\ }\href@noop {} {\bibfield  {journal} {\bibinfo  {journal} {Prog.
  Theor. Phys.}\ }\textbf {\bibinfo {volume} {46}},\ \bibinfo {pages} {1337}
  (\bibinfo {year} {1971})}\BibitemShut {NoStop}%
\bibitem [{\citenamefont {Suzuki}(1976)}]{Suzuki1976}%
  \BibitemOpen
  \bibfield  {author} {\bibinfo {author} {\bibfnamefont {M.}~\bibnamefont
  {Suzuki}},\ }\href@noop {} {\bibfield  {journal} {\bibinfo  {journal} {Prog.
  of Theor. Phys.}\ }\textbf {\bibinfo {volume} {56}},\ \bibinfo {pages} {1454}
  (\bibinfo {year} {1976})}\BibitemShut {NoStop}%
\bibitem [{\citenamefont {Jullien}\ \emph {et~al.}(1978)\citenamefont
  {Jullien}, \citenamefont {Pfeuty}, \citenamefont {Fields},\ and\
  \citenamefont {Doniach}}]{Jullien1978}%
  \BibitemOpen
  \bibfield  {author} {\bibinfo {author} {\bibfnamefont {R.}~\bibnamefont
  {Jullien}}, \bibinfo {author} {\bibfnamefont {P.}~\bibnamefont {Pfeuty}},
  \bibinfo {author} {\bibfnamefont {J.~N.}\ \bibnamefont {Fields}}, \ and\
  \bibinfo {author} {\bibfnamefont {S.}~\bibnamefont {Doniach}},\ }\href
  {\doibase 10.1103/PhysRevB.18.3568} {\bibfield  {journal} {\bibinfo
  {journal} {Phys. Rev. B}\ }\textbf {\bibinfo {volume} {18}},\ \bibinfo
  {pages} {3568} (\bibinfo {year} {1978})}\BibitemShut {NoStop}%
\bibitem [{\citenamefont {Kopp}\ and\ \citenamefont
  {Chakravarty}(2005)}]{Chakravarty2005}%
  \BibitemOpen
  \bibfield  {author} {\bibinfo {author} {\bibfnamefont {A.}~\bibnamefont
  {Kopp}}\ and\ \bibinfo {author} {\bibfnamefont {S.}~\bibnamefont
  {Chakravarty}},\ }\href@noop {} {\bibfield  {journal} {\bibinfo  {journal}
  {Nat. Phys.}\ }\textbf {\bibinfo {volume} {1}},\ \bibinfo {pages} {53}
  (\bibinfo {year} {2005})}\BibitemShut {NoStop}%
\bibitem [{\citenamefont {Wu}\ \emph {et~al.}(2018)\citenamefont {Wu},
  \citenamefont {Zhu},\ and\ \citenamefont {Si}}]{WuPRB2018}%
  \BibitemOpen
  \bibfield  {author} {\bibinfo {author} {\bibfnamefont {J.}~\bibnamefont
  {Wu}}, \bibinfo {author} {\bibfnamefont {L.}~\bibnamefont {Zhu}}, \ and\
  \bibinfo {author} {\bibfnamefont {Q.}~\bibnamefont {Si}},\ }\href {\doibase
  10.1103/PhysRevB.97.245127} {\bibfield  {journal} {\bibinfo  {journal} {Phys.
  Rev. B}\ }\textbf {\bibinfo {volume} {97}},\ \bibinfo {pages} {245127}
  (\bibinfo {year} {2018})}\BibitemShut {NoStop}%
\bibitem [{\citenamefont {Wang}\ \emph
  {et~al.}(2018{\natexlab{a}})\citenamefont {Wang}, \citenamefont {Lorenz},
  \citenamefont {Gorbunov}, \citenamefont {Cong}, \citenamefont {Kohama},
  \citenamefont {Niesen}, \citenamefont {Breunig}, \citenamefont {Engelmayer},
  \citenamefont {Herman}, \citenamefont {Wu}, \citenamefont {Kindo},
  \citenamefont {Wosnitza}, \citenamefont {Zherlitsyn},\ and\ \citenamefont
  {Loidl}}]{WangPRL2018}%
  \BibitemOpen
  \bibfield  {author} {\bibinfo {author} {\bibfnamefont {Z.}~\bibnamefont
  {Wang}}, \bibinfo {author} {\bibfnamefont {T.}~\bibnamefont {Lorenz}},
  \bibinfo {author} {\bibfnamefont {D.~I.}\ \bibnamefont {Gorbunov}}, \bibinfo
  {author} {\bibfnamefont {P.~T.}\ \bibnamefont {Cong}}, \bibinfo {author}
  {\bibfnamefont {Y.}~\bibnamefont {Kohama}}, \bibinfo {author} {\bibfnamefont
  {S.}~\bibnamefont {Niesen}}, \bibinfo {author} {\bibfnamefont
  {O.}~\bibnamefont {Breunig}}, \bibinfo {author} {\bibfnamefont
  {J.}~\bibnamefont {Engelmayer}}, \bibinfo {author} {\bibfnamefont
  {A.}~\bibnamefont {Herman}}, \bibinfo {author} {\bibfnamefont
  {J.}~\bibnamefont {Wu}}, \bibinfo {author} {\bibfnamefont {K.}~\bibnamefont
  {Kindo}}, \bibinfo {author} {\bibfnamefont {J.}~\bibnamefont {Wosnitza}},
  \bibinfo {author} {\bibfnamefont {S.}~\bibnamefont {Zherlitsyn}}, \ and\
  \bibinfo {author} {\bibfnamefont {A.}~\bibnamefont {Loidl}},\ }\href
  {\doibase 10.1103/PhysRevLett.120.207205} {\bibfield  {journal} {\bibinfo
  {journal} {Phys. Rev. Lett.}\ }\textbf {\bibinfo {volume} {120}},\ \bibinfo
  {pages} {207205} (\bibinfo {year} {2018}{\natexlab{a}})}\BibitemShut
  {NoStop}%
\bibitem [{\citenamefont {Belavin}\ \emph {et~al.}(1984)\citenamefont
  {Belavin}, \citenamefont {Polyakov},\ and\ \citenamefont
  {Zamolodchikov}}]{Belavin1984}%
  \BibitemOpen
  \bibfield  {author} {\bibinfo {author} {\bibfnamefont {A.}~\bibnamefont
  {Belavin}}, \bibinfo {author} {\bibfnamefont {A.}~\bibnamefont {Polyakov}}, \
  and\ \bibinfo {author} {\bibfnamefont {A.}~\bibnamefont {Zamolodchikov}},\
  }\href@noop {} {\bibfield  {journal} {\bibinfo  {journal} {Nucl. Phys. B}\
  }\textbf {\bibinfo {volume} {241}},\ \bibinfo {pages} {333} (\bibinfo {year}
  {1984})}\BibitemShut {NoStop}%
\bibitem [{\citenamefont {Coldea}\ \emph {et~al.}(2010)\citenamefont {Coldea},
  \citenamefont {Tennant}, \citenamefont {Wheeler}, \citenamefont {Wawrzynska},
  \citenamefont {Prabhakaran}, \citenamefont {Telling}, \citenamefont
  {Habicht}, \citenamefont {Smeibidl},\ and\ \citenamefont
  {Kiefer}}]{ColdeaE8Science2010}%
  \BibitemOpen
  \bibfield  {author} {\bibinfo {author} {\bibfnamefont {R.}~\bibnamefont
  {Coldea}}, \bibinfo {author} {\bibfnamefont {D.~A.}\ \bibnamefont {Tennant}},
  \bibinfo {author} {\bibfnamefont {E.~M.}\ \bibnamefont {Wheeler}}, \bibinfo
  {author} {\bibfnamefont {E.}~\bibnamefont {Wawrzynska}}, \bibinfo {author}
  {\bibfnamefont {D.}~\bibnamefont {Prabhakaran}}, \bibinfo {author}
  {\bibfnamefont {M.}~\bibnamefont {Telling}}, \bibinfo {author} {\bibfnamefont
  {K.}~\bibnamefont {Habicht}}, \bibinfo {author} {\bibfnamefont
  {P.}~\bibnamefont {Smeibidl}}, \ and\ \bibinfo {author} {\bibfnamefont
  {K.}~\bibnamefont {Kiefer}},\ }\href {\doibase 10.1126/science.1180085}
  {\bibfield  {journal} {\bibinfo  {journal} {Science}\ }\textbf {\bibinfo
  {volume} {327}},\ \bibinfo {pages} {177} (\bibinfo {year}
  {2010})}\BibitemShut {NoStop}%
\bibitem [{\citenamefont {He}\ \emph {et~al.}(2006)\citenamefont {He},
  \citenamefont {Taniyama},\ and\ \citenamefont {Itoh}}]{HePRB2006SCVO}%
  \BibitemOpen
  \bibfield  {author} {\bibinfo {author} {\bibfnamefont {Z.}~\bibnamefont
  {He}}, \bibinfo {author} {\bibfnamefont {T.}~\bibnamefont {Taniyama}}, \ and\
  \bibinfo {author} {\bibfnamefont {M.}~\bibnamefont {Itoh}},\ }\href {\doibase
  10.1103/PhysRevB.73.212406} {\bibfield  {journal} {\bibinfo  {journal} {Phys.
  Rev. B}\ }\textbf {\bibinfo {volume} {73}},\ \bibinfo {pages} {212406}
  (\bibinfo {year} {2006})}\BibitemShut {NoStop}%
\bibitem [{\citenamefont {Bera}\ \emph {et~al.}(2017)\citenamefont {Bera},
  \citenamefont {Lake}, \citenamefont {Essler}, \citenamefont {Vanderstraeten},
  \citenamefont {Hubig}, \citenamefont {Schollw\"ock}, \citenamefont {Islam},
  \citenamefont {Schneidewind},\ and\ \citenamefont
  {Quintero-Castro}}]{BeraPRB2017}%
  \BibitemOpen
  \bibfield  {author} {\bibinfo {author} {\bibfnamefont {A.~K.}\ \bibnamefont
  {Bera}}, \bibinfo {author} {\bibfnamefont {B.}~\bibnamefont {Lake}}, \bibinfo
  {author} {\bibfnamefont {F.~H.~L.}\ \bibnamefont {Essler}}, \bibinfo {author}
  {\bibfnamefont {L.}~\bibnamefont {Vanderstraeten}}, \bibinfo {author}
  {\bibfnamefont {C.}~\bibnamefont {Hubig}}, \bibinfo {author} {\bibfnamefont
  {U.}~\bibnamefont {Schollw\"ock}}, \bibinfo {author} {\bibfnamefont {A.~T.
  M.~N.}\ \bibnamefont {Islam}}, \bibinfo {author} {\bibfnamefont
  {A.}~\bibnamefont {Schneidewind}}, \ and\ \bibinfo {author} {\bibfnamefont
  {D.~L.}\ \bibnamefont {Quintero-Castro}},\ }\href {\doibase
  10.1103/PhysRevB.96.054423} {\bibfield  {journal} {\bibinfo  {journal} {Phys.
  Rev. B}\ }\textbf {\bibinfo {volume} {96}},\ \bibinfo {pages} {054423}
  (\bibinfo {year} {2017})}\BibitemShut {NoStop}%
\bibitem [{\citenamefont {Bera}\ \emph {et~al.}(2014)\citenamefont {Bera},
  \citenamefont {Lake}, \citenamefont {Stein},\ and\ \citenamefont
  {Zander}}]{BeraPRB2014}%
  \BibitemOpen
  \bibfield  {author} {\bibinfo {author} {\bibfnamefont {A.~K.}\ \bibnamefont
  {Bera}}, \bibinfo {author} {\bibfnamefont {B.}~\bibnamefont {Lake}}, \bibinfo
  {author} {\bibfnamefont {W.-D.}\ \bibnamefont {Stein}}, \ and\ \bibinfo
  {author} {\bibfnamefont {S.}~\bibnamefont {Zander}},\ }\href {\doibase
  10.1103/PhysRevB.89.094402} {\bibfield  {journal} {\bibinfo  {journal} {Phys.
  Rev. B}\ }\textbf {\bibinfo {volume} {89}},\ \bibinfo {pages} {094402}
  (\bibinfo {year} {2014})}\BibitemShut {NoStop}%
\bibitem [{\citenamefont {Wang}\ \emph
  {et~al.}(2018{\natexlab{b}})\citenamefont {Wang}, \citenamefont {Wu},
  \citenamefont {Yang}, \citenamefont {Bera}, \citenamefont {Kamenskyi},
  \citenamefont {Islam}, \citenamefont {Xu}, \citenamefont {Law}, \citenamefont
  {Lake}, \citenamefont {Wu},\ and\ \citenamefont {Loidl}}]{WangNature2018}%
  \BibitemOpen
  \bibfield  {author} {\bibinfo {author} {\bibfnamefont {Z.}~\bibnamefont
  {Wang}}, \bibinfo {author} {\bibfnamefont {J.}~\bibnamefont {Wu}}, \bibinfo
  {author} {\bibfnamefont {W.}~\bibnamefont {Yang}}, \bibinfo {author}
  {\bibfnamefont {A.~K.}\ \bibnamefont {Bera}}, \bibinfo {author}
  {\bibfnamefont {D.}~\bibnamefont {Kamenskyi}}, \bibinfo {author}
  {\bibfnamefont {A.~T. M.~N.}\ \bibnamefont {Islam}}, \bibinfo {author}
  {\bibfnamefont {S.}~\bibnamefont {Xu}}, \bibinfo {author} {\bibfnamefont
  {J.~M.}\ \bibnamefont {Law}}, \bibinfo {author} {\bibfnamefont
  {B.}~\bibnamefont {Lake}}, \bibinfo {author} {\bibfnamefont {C.}~\bibnamefont
  {Wu}}, \ and\ \bibinfo {author} {\bibfnamefont {A.}~\bibnamefont {Loidl}},\
  }\href@noop {} {\bibfield  {journal} {\bibinfo  {journal} {Nature}\ }\textbf
  {\bibinfo {volume} {554}},\ \bibinfo {pages} {219} (\bibinfo {year}
  {2018}{\natexlab{b}})}\BibitemShut {NoStop}%
\bibitem [{\citenamefont {Faure}\ \emph {et~al.}(2018)\citenamefont {Faure},
  \citenamefont {Takayoshi}, \citenamefont {Petit}, \citenamefont {Simonet},
  \citenamefont {Raymond}, \citenamefont {Regnault}, \citenamefont {Boehm},
  \citenamefont {White}, \citenamefont {M{\aa}nsson}, \citenamefont
  {R{\"u}egg}, \citenamefont {Lejay}, \citenamefont {Canals}, \citenamefont
  {Lorenz}, \citenamefont {Furuya}, \citenamefont {Giamarchi},\ and\
  \citenamefont {Grenier}}]{NP2018Faure}%
  \BibitemOpen
  \bibfield  {author} {\bibinfo {author} {\bibfnamefont {Q.}~\bibnamefont
  {Faure}}, \bibinfo {author} {\bibfnamefont {S.}~\bibnamefont {Takayoshi}},
  \bibinfo {author} {\bibfnamefont {S.}~\bibnamefont {Petit}}, \bibinfo
  {author} {\bibfnamefont {V.}~\bibnamefont {Simonet}}, \bibinfo {author}
  {\bibfnamefont {S.}~\bibnamefont {Raymond}}, \bibinfo {author} {\bibfnamefont
  {L.-P.}\ \bibnamefont {Regnault}}, \bibinfo {author} {\bibfnamefont
  {M.}~\bibnamefont {Boehm}}, \bibinfo {author} {\bibfnamefont {J.~S.}\
  \bibnamefont {White}}, \bibinfo {author} {\bibfnamefont {M.}~\bibnamefont
  {M{\aa}nsson}}, \bibinfo {author} {\bibfnamefont {C.}~\bibnamefont
  {R{\"u}egg}}, \bibinfo {author} {\bibfnamefont {P.}~\bibnamefont {Lejay}},
  \bibinfo {author} {\bibfnamefont {B.}~\bibnamefont {Canals}}, \bibinfo
  {author} {\bibfnamefont {T.}~\bibnamefont {Lorenz}}, \bibinfo {author}
  {\bibfnamefont {S.~C.}\ \bibnamefont {Furuya}}, \bibinfo {author}
  {\bibfnamefont {T.}~\bibnamefont {Giamarchi}}, \ and\ \bibinfo {author}
  {\bibfnamefont {B.}~\bibnamefont {Grenier}},\ }\href {\doibase
  10.1038/s41567-018-0126-8} {\bibfield  {journal} {\bibinfo  {journal} {Nature
  Physics}\ }\textbf {\bibinfo {volume} {14}},\ \bibinfo {pages} {716}
  (\bibinfo {year} {2018})}\BibitemShut {NoStop}%
\bibitem [{\citenamefont {He}\ \emph {et~al.}(2005)\citenamefont {He},
  \citenamefont {Taniyama}, \citenamefont {Ky\^omen},\ and\ \citenamefont
  {Itoh}}]{HePRB2005BCVO}%
  \BibitemOpen
  \bibfield  {author} {\bibinfo {author} {\bibfnamefont {Z.}~\bibnamefont
  {He}}, \bibinfo {author} {\bibfnamefont {T.}~\bibnamefont {Taniyama}},
  \bibinfo {author} {\bibfnamefont {T.}~\bibnamefont {Ky\^omen}}, \ and\
  \bibinfo {author} {\bibfnamefont {M.}~\bibnamefont {Itoh}},\ }\href {\doibase
  10.1103/PhysRevB.72.172403} {\bibfield  {journal} {\bibinfo  {journal} {Phys.
  Rev. B}\ }\textbf {\bibinfo {volume} {72}},\ \bibinfo {pages} {172403}
  (\bibinfo {year} {2005})}\BibitemShut {NoStop}%
\bibitem [{\citenamefont {Kimura}\ \emph {et~al.}(2007)\citenamefont {Kimura},
  \citenamefont {Yashiro}, \citenamefont {Okunishi}, \citenamefont {Hagiwara},
  \citenamefont {He}, \citenamefont {Kindo}, \citenamefont {Taniyama},\ and\
  \citenamefont {Itoh}}]{KimuraPRL2007}%
  \BibitemOpen
  \bibfield  {author} {\bibinfo {author} {\bibfnamefont {S.}~\bibnamefont
  {Kimura}}, \bibinfo {author} {\bibfnamefont {H.}~\bibnamefont {Yashiro}},
  \bibinfo {author} {\bibfnamefont {K.}~\bibnamefont {Okunishi}}, \bibinfo
  {author} {\bibfnamefont {M.}~\bibnamefont {Hagiwara}}, \bibinfo {author}
  {\bibfnamefont {Z.}~\bibnamefont {He}}, \bibinfo {author} {\bibfnamefont
  {K.}~\bibnamefont {Kindo}}, \bibinfo {author} {\bibfnamefont
  {T.}~\bibnamefont {Taniyama}}, \ and\ \bibinfo {author} {\bibfnamefont
  {M.}~\bibnamefont {Itoh}},\ }\href {\doibase 10.1103/PhysRevLett.99.087602}
  {\bibfield  {journal} {\bibinfo  {journal} {Phys. Rev. Lett.}\ }\textbf
  {\bibinfo {volume} {99}},\ \bibinfo {pages} {087602} (\bibinfo {year}
  {2007})}\BibitemShut {NoStop}%
\bibitem [{\citenamefont {Lake}\ \emph {et~al.}(2013)\citenamefont {Lake},
  \citenamefont {Tennant}, \citenamefont {Caux}, \citenamefont {Barthel},
  \citenamefont {Schollw\"ock}, \citenamefont {Nagler},\ and\ \citenamefont
  {Frost}}]{LakePRL2013}%
  \BibitemOpen
  \bibfield  {author} {\bibinfo {author} {\bibfnamefont {B.}~\bibnamefont
  {Lake}}, \bibinfo {author} {\bibfnamefont {D.~A.}\ \bibnamefont {Tennant}},
  \bibinfo {author} {\bibfnamefont {J.-S.}\ \bibnamefont {Caux}}, \bibinfo
  {author} {\bibfnamefont {T.}~\bibnamefont {Barthel}}, \bibinfo {author}
  {\bibfnamefont {U.}~\bibnamefont {Schollw\"ock}}, \bibinfo {author}
  {\bibfnamefont {S.~E.}\ \bibnamefont {Nagler}}, \ and\ \bibinfo {author}
  {\bibfnamefont {C.~D.}\ \bibnamefont {Frost}},\ }\href {\doibase
  10.1103/PhysRevLett.111.137205} {\bibfield  {journal} {\bibinfo  {journal}
  {Phys. Rev. Lett.}\ }\textbf {\bibinfo {volume} {111}},\ \bibinfo {pages}
  {137205} (\bibinfo {year} {2013})}\BibitemShut {NoStop}%
\bibitem [{\citenamefont {Klanj\ifmmode~\check{s}\else \v{s}\fi{}ek}\ \emph
  {et~al.}(2015)\citenamefont {Klanj\ifmmode~\check{s}\else \v{s}\fi{}ek},
  \citenamefont {Horvati\ifmmode~\acute{c}\else \'{c}\fi{}}, \citenamefont
  {Kr\"amer}, \citenamefont {Mukhopadhyay}, \citenamefont {Mayaffre},
  \citenamefont {Berthier}, \citenamefont {Can\'evet}, \citenamefont {Grenier},
  \citenamefont {Lejay},\ and\ \citenamefont {Orignac}}]{KlanjsekPRB2015}%
  \BibitemOpen
  \bibfield  {author} {\bibinfo {author} {\bibfnamefont {M.}~\bibnamefont
  {Klanj\ifmmode~\check{s}\else \v{s}\fi{}ek}}, \bibinfo {author}
  {\bibfnamefont {M.}~\bibnamefont {Horvati\ifmmode~\acute{c}\else
  \'{c}\fi{}}}, \bibinfo {author} {\bibfnamefont {S.}~\bibnamefont {Kr\"amer}},
  \bibinfo {author} {\bibfnamefont {S.}~\bibnamefont {Mukhopadhyay}}, \bibinfo
  {author} {\bibfnamefont {H.}~\bibnamefont {Mayaffre}}, \bibinfo {author}
  {\bibfnamefont {C.}~\bibnamefont {Berthier}}, \bibinfo {author}
  {\bibfnamefont {E.}~\bibnamefont {Can\'evet}}, \bibinfo {author}
  {\bibfnamefont {B.}~\bibnamefont {Grenier}}, \bibinfo {author} {\bibfnamefont
  {P.}~\bibnamefont {Lejay}}, \ and\ \bibinfo {author} {\bibfnamefont
  {E.}~\bibnamefont {Orignac}},\ }\href {\doibase 10.1103/PhysRevB.92.060408}
  {\bibfield  {journal} {\bibinfo  {journal} {Phys. Rev. B}\ }\textbf {\bibinfo
  {volume} {92}},\ \bibinfo {pages} {060408} (\bibinfo {year}
  {2015})}\BibitemShut {NoStop}%
\bibitem [{\citenamefont {Kimura}\ \emph {et~al.}(2013)\citenamefont {Kimura},
  \citenamefont {Okunishi}, \citenamefont {Hagiwara}, \citenamefont {Kindo},
  \citenamefont {He}, \citenamefont {Taniyama}, \citenamefont {Itoh},
  \citenamefont {Koyama},\ and\ \citenamefont {Watanabe}}]{KimuraJPSJ2013}%
  \BibitemOpen
  \bibfield  {author} {\bibinfo {author} {\bibfnamefont {S.}~\bibnamefont
  {Kimura}}, \bibinfo {author} {\bibfnamefont {K.}~\bibnamefont {Okunishi}},
  \bibinfo {author} {\bibfnamefont {M.}~\bibnamefont {Hagiwara}}, \bibinfo
  {author} {\bibfnamefont {K.}~\bibnamefont {Kindo}}, \bibinfo {author}
  {\bibfnamefont {Z.}~\bibnamefont {He}}, \bibinfo {author} {\bibfnamefont
  {T.}~\bibnamefont {Taniyama}}, \bibinfo {author} {\bibfnamefont
  {M.}~\bibnamefont {Itoh}}, \bibinfo {author} {\bibfnamefont {K.}~\bibnamefont
  {Koyama}}, \ and\ \bibinfo {author} {\bibfnamefont {K.}~\bibnamefont
  {Watanabe}},\ }\href {\doibase 10.7566/JPSJ.82.033706} {\bibfield  {journal}
  {\bibinfo  {journal} {Journal of the Physical Society of Japan}\ }\textbf
  {\bibinfo {volume} {82}},\ \bibinfo {pages} {033706} (\bibinfo {year}
  {2013})}\BibitemShut {NoStop}%
\bibitem [{\citenamefont {Cui}\ \emph {et~al.}()\citenamefont {Cui},
  \citenamefont {Zou}, \citenamefont {Xi}, \citenamefont {He}, \citenamefont
  {Yang}, \citenamefont {Shu}, \citenamefont {Zhang}, \citenamefont {Hu},
  \citenamefont {Chen}, \citenamefont {Yu}, \citenamefont {Wu},\ and\
  \citenamefont {Yu}}]{YuRecent}%
  \BibitemOpen
  \bibfield  {author} {\bibinfo {author} {\bibfnamefont {Y.}~\bibnamefont
  {Cui}}, \bibinfo {author} {\bibfnamefont {H.}~\bibnamefont {Zou}}, \bibinfo
  {author} {\bibfnamefont {N.}~\bibnamefont {Xi}}, \bibinfo {author}
  {\bibfnamefont {Z.}~\bibnamefont {He}}, \bibinfo {author} {\bibfnamefont
  {Y.~X.}\ \bibnamefont {Yang}}, \bibinfo {author} {\bibfnamefont
  {L.}~\bibnamefont {Shu}}, \bibinfo {author} {\bibfnamefont {G.}~\bibnamefont
  {Zhang}}, \bibinfo {author} {\bibfnamefont {Z.}~\bibnamefont {Hu}}, \bibinfo
  {author} {\bibfnamefont {T.}~\bibnamefont {Chen}}, \bibinfo {author}
  {\bibfnamefont {R.}~\bibnamefont {Yu}}, \bibinfo {author} {\bibfnamefont
  {J.}~\bibnamefont {Wu}}, \ and\ \bibinfo {author} {\bibfnamefont
  {W.}~\bibnamefont {Yu}},\ }\href@noop {} {}\Eprint
  {http://arxiv.org/abs/submitted, 2019} {submitted, 2019} \BibitemShut
  {NoStop}%
\bibitem [{\citenamefont {Niesen}\ \emph {et~al.}(2013)\citenamefont {Niesen},
  \citenamefont {Kolland}, \citenamefont {Seher}, \citenamefont {Breunig},
  \citenamefont {Valldor}, \citenamefont {Braden}, \citenamefont {Grenier},\
  and\ \citenamefont {Lorenz}}]{NiesenPRB2013}%
  \BibitemOpen
  \bibfield  {author} {\bibinfo {author} {\bibfnamefont {S.~K.}\ \bibnamefont
  {Niesen}}, \bibinfo {author} {\bibfnamefont {G.}~\bibnamefont {Kolland}},
  \bibinfo {author} {\bibfnamefont {M.}~\bibnamefont {Seher}}, \bibinfo
  {author} {\bibfnamefont {O.}~\bibnamefont {Breunig}}, \bibinfo {author}
  {\bibfnamefont {M.}~\bibnamefont {Valldor}}, \bibinfo {author} {\bibfnamefont
  {M.}~\bibnamefont {Braden}}, \bibinfo {author} {\bibfnamefont
  {B.}~\bibnamefont {Grenier}}, \ and\ \bibinfo {author} {\bibfnamefont
  {T.}~\bibnamefont {Lorenz}},\ }\href {\doibase 10.1103/PhysRevB.87.224413}
  {\bibfield  {journal} {\bibinfo  {journal} {Phys. Rev. B}\ }\textbf {\bibinfo
  {volume} {87}},\ \bibinfo {pages} {224413} (\bibinfo {year}
  {2013})}\BibitemShut {NoStop}%
\bibitem [{\citenamefont {Wang}\ \emph {et~al.}(2016)\citenamefont {Wang},
  \citenamefont {Wu}, \citenamefont {Xu}, \citenamefont {Yang}, \citenamefont
  {Wu}, \citenamefont {Bera}, \citenamefont {Islam}, \citenamefont {Lake},
  \citenamefont {Kamenskyi}, \citenamefont {Gogoi}, \citenamefont {Engelkamp},
  \citenamefont {Wang}, \citenamefont {Deisenhofer},\ and\ \citenamefont
  {Loidl}}]{WangPRB2016}%
  \BibitemOpen
  \bibfield  {author} {\bibinfo {author} {\bibfnamefont {Z.}~\bibnamefont
  {Wang}}, \bibinfo {author} {\bibfnamefont {J.}~\bibnamefont {Wu}}, \bibinfo
  {author} {\bibfnamefont {S.}~\bibnamefont {Xu}}, \bibinfo {author}
  {\bibfnamefont {W.}~\bibnamefont {Yang}}, \bibinfo {author} {\bibfnamefont
  {C.}~\bibnamefont {Wu}}, \bibinfo {author} {\bibfnamefont {A.~K.}\
  \bibnamefont {Bera}}, \bibinfo {author} {\bibfnamefont {A.~T. M.~N.}\
  \bibnamefont {Islam}}, \bibinfo {author} {\bibfnamefont {B.}~\bibnamefont
  {Lake}}, \bibinfo {author} {\bibfnamefont {D.}~\bibnamefont {Kamenskyi}},
  \bibinfo {author} {\bibfnamefont {P.}~\bibnamefont {Gogoi}}, \bibinfo
  {author} {\bibfnamefont {H.}~\bibnamefont {Engelkamp}}, \bibinfo {author}
  {\bibfnamefont {N.}~\bibnamefont {Wang}}, \bibinfo {author} {\bibfnamefont
  {J.}~\bibnamefont {Deisenhofer}}, \ and\ \bibinfo {author} {\bibfnamefont
  {A.}~\bibnamefont {Loidl}},\ }\href {\doibase 10.1103/PhysRevB.94.125130}
  {\bibfield  {journal} {\bibinfo  {journal} {Phys. Rev. B}\ }\textbf {\bibinfo
  {volume} {94}},\ \bibinfo {pages} {125130} (\bibinfo {year}
  {2016})}\BibitemShut {NoStop}%
\bibitem [{\citenamefont {Vidal}(2007)}]{iTEBD1}%
  \BibitemOpen
  \bibfield  {author} {\bibinfo {author} {\bibfnamefont {G.}~\bibnamefont
  {Vidal}},\ }\href {\doibase 10.1103/PhysRevLett.98.070201} {\bibfield
  {journal} {\bibinfo  {journal} {Phys. Rev. Lett.}\ }\textbf {\bibinfo
  {volume} {98}},\ \bibinfo {pages} {070201} (\bibinfo {year}
  {2007})}\BibitemShut {NoStop}%
\bibitem [{\citenamefont {Or\'us}\ and\ \citenamefont {Vidal}(2008)}]{iTEBD2}%
  \BibitemOpen
  \bibfield  {author} {\bibinfo {author} {\bibfnamefont {R.}~\bibnamefont
  {Or\'us}}\ and\ \bibinfo {author} {\bibfnamefont {G.}~\bibnamefont {Vidal}},\
  }\href {\doibase 10.1103/PhysRevB.78.155117} {\bibfield  {journal} {\bibinfo
  {journal} {Phys. Rev. B}\ }\textbf {\bibinfo {volume} {78}},\ \bibinfo
  {pages} {155117} (\bibinfo {year} {2008})}\BibitemShut {NoStop}%
\bibitem [{\citenamefont {Wang}\ \emph {et~al.}(2015)\citenamefont {Wang},
  \citenamefont {Schmidt}, \citenamefont {Bera}, \citenamefont {Islam},
  \citenamefont {Lake}, \citenamefont {Loidl},\ and\ \citenamefont
  {Deisenhofer}}]{WangPRB2015}%
  \BibitemOpen
  \bibfield  {author} {\bibinfo {author} {\bibfnamefont {Z.}~\bibnamefont
  {Wang}}, \bibinfo {author} {\bibfnamefont {M.}~\bibnamefont {Schmidt}},
  \bibinfo {author} {\bibfnamefont {A.~K.}\ \bibnamefont {Bera}}, \bibinfo
  {author} {\bibfnamefont {A.~T. M.~N.}\ \bibnamefont {Islam}}, \bibinfo
  {author} {\bibfnamefont {B.}~\bibnamefont {Lake}}, \bibinfo {author}
  {\bibfnamefont {A.}~\bibnamefont {Loidl}}, \ and\ \bibinfo {author}
  {\bibfnamefont {J.}~\bibnamefont {Deisenhofer}},\ }\href {\doibase
  10.1103/PhysRevB.91.140404} {\bibfield  {journal} {\bibinfo  {journal} {Phys.
  Rev. B}\ }\textbf {\bibinfo {volume} {91}},\ \bibinfo {pages} {140404}
  (\bibinfo {year} {2015})}\BibitemShut {NoStop}%
\bibitem [{\citenamefont {Calabrese}\ and\ \citenamefont
  {Cardy}(2004)}]{Cardy2004}%
  \BibitemOpen
  \bibfield  {author} {\bibinfo {author} {\bibfnamefont {P.}~\bibnamefont
  {Calabrese}}\ and\ \bibinfo {author} {\bibfnamefont {J.}~\bibnamefont
  {Cardy}},\ }\href@noop {} {\bibfield  {journal} {\bibinfo  {journal} {Journal
  of Statistical Mechanics: Theory and Experiment}\ }\textbf {\bibinfo {volume}
  {2004}},\ \bibinfo {pages} {P06002} (\bibinfo {year} {2004})}\BibitemShut
  {NoStop}%
\bibitem [{\citenamefont {Takayoshi}\ \emph {et~al.}(2018)\citenamefont
  {Takayoshi}, \citenamefont {Furuya},\ and\ \citenamefont
  {Giamarchi}}]{Giamarchi2018}%
  \BibitemOpen
  \bibfield  {author} {\bibinfo {author} {\bibfnamefont {S.}~\bibnamefont
  {Takayoshi}}, \bibinfo {author} {\bibfnamefont {S.~C.}\ \bibnamefont
  {Furuya}}, \ and\ \bibinfo {author} {\bibfnamefont {T.}~\bibnamefont
  {Giamarchi}},\ }\href {\doibase 10.1103/PhysRevB.98.184429} {\bibfield
  {journal} {\bibinfo  {journal} {Phys. Rev. B}\ }\textbf {\bibinfo {volume}
  {98}},\ \bibinfo {pages} {184429} (\bibinfo {year} {2018})}\BibitemShut
  {NoStop}%
\bibitem [{\citenamefont {Hieida}\ \emph {et~al.}(2001)\citenamefont {Hieida},
  \citenamefont {Okunishi},\ and\ \citenamefont {Akutsu}}]{YasuhiroPRB2001}%
  \BibitemOpen
  \bibfield  {author} {\bibinfo {author} {\bibfnamefont {Y.}~\bibnamefont
  {Hieida}}, \bibinfo {author} {\bibfnamefont {K.}~\bibnamefont {Okunishi}}, \
  and\ \bibinfo {author} {\bibfnamefont {Y.}~\bibnamefont {Akutsu}},\ }\href
  {\doibase 10.1103/PhysRevB.64.224422} {\bibfield  {journal} {\bibinfo
  {journal} {Phys. Rev. B}\ }\textbf {\bibinfo {volume} {64}},\ \bibinfo
  {pages} {224422} (\bibinfo {year} {2001})}\BibitemShut {NoStop}%
\bibitem [{\citenamefont {Dmitriev}\ \emph {et~al.}(2002)\citenamefont
  {Dmitriev}, \citenamefont {Krivnov}, \citenamefont {Ovchinnikov},\ and\
  \citenamefont {Langari}}]{Dmitriev2002}%
  \BibitemOpen
  \bibfield  {author} {\bibinfo {author} {\bibfnamefont {D.~V.}\ \bibnamefont
  {Dmitriev}}, \bibinfo {author} {\bibfnamefont {V.~Y.}\ \bibnamefont
  {Krivnov}}, \bibinfo {author} {\bibfnamefont {A.~A.}\ \bibnamefont
  {Ovchinnikov}}, \ and\ \bibinfo {author} {\bibfnamefont {A.}~\bibnamefont
  {Langari}},\ }\href {\doibase 10.1134/1.1513828} {\bibfield  {journal}
  {\bibinfo  {journal} {Journal of Experimental and Theoretical Physics}\
  }\textbf {\bibinfo {volume} {95}},\ \bibinfo {pages} {538} (\bibinfo {year}
  {2002})}\BibitemShut {NoStop}%
\bibitem [{\citenamefont {Caux}\ \emph {et~al.}(2003)\citenamefont {Caux},
  \citenamefont {Essler},\ and\ \citenamefont {L\"ow}}]{CauxPRB2003}%
  \BibitemOpen
  \bibfield  {author} {\bibinfo {author} {\bibfnamefont {J.-S.}\ \bibnamefont
  {Caux}}, \bibinfo {author} {\bibfnamefont {F.~H.~L.}\ \bibnamefont {Essler}},
  \ and\ \bibinfo {author} {\bibfnamefont {U.}~\bibnamefont {L\"ow}},\ }\href
  {\doibase 10.1103/PhysRevB.68.134431} {\bibfield  {journal} {\bibinfo
  {journal} {Phys. Rev. B}\ }\textbf {\bibinfo {volume} {68}},\ \bibinfo
  {pages} {134431} (\bibinfo {year} {2003})}\BibitemShut {NoStop}%
\bibitem [{\citenamefont {Zamolodchikov}(1989)}]{E8_1989}%
  \BibitemOpen
  \bibfield  {author} {\bibinfo {author} {\bibfnamefont {A.}~\bibnamefont
  {Zamolodchikov}},\ }\href@noop {} {\bibfield  {journal} {\bibinfo  {journal}
  {Int. J. Mod. Phys. A}\ }\textbf {\bibinfo {volume} {4}},\ \bibinfo {pages}
  {4235} (\bibinfo {year} {1989})}\BibitemShut {NoStop}%
\end{thebibliography}%
\end{document}